\documentclass[12pt]{article}
\pdfoutput=1

\usepackage{jheppub}
\usepackage{amssymb,amsfonts,amsmath,amsthm}
\usepackage{mathrsfs}
\usepackage{bbold}
\usepackage{tikz}
\usepackage{caption}
\usepackage{subcaption}
\usetikzlibrary{shapes.geometric,calc}
\usetikzlibrary{shapes,arrows,chains}
\usetikzlibrary{decorations.markings}
\usetikzlibrary{decorations.pathmorphing}
\usetikzlibrary{shapes.multipart}
\tikzset{snake it/.style={decorate, decoration=snake}}

\newcommand{\be}{\begin{equation}}
\newcommand{\ee}{\end{equation}}
\newcommand{\bea}{\begin{eqnarray}}
\newcommand{\eea}{\end{eqnarray}}

\newcommand{\CA}{\mathcal{A}}

\newcommand{\CC}{\mathcal{C}}

\newcommand{\CI}{\mathcal{I}}

\newcommand{\CL}{\mathcal{L}}
\newcommand{\CN}{\mathcal{N}}
\newcommand{\CM}{\mathcal{M}}

\newcommand{\CV}{\mathcal{V}}

\newcommand{\lr}{\left (}
\newcommand{\rr}{\right )}

\newcommand\qt\tau

\newcommand{\p}{\partial}

\renewcommand{\tilde}[1]{\widetilde{#1}}

\makeatletter
\renewcommand{\@seccntformat}[1]{\csname the#1\endcsname.\,\,}
\makeatother
\allowdisplaybreaks

\definecolor{darkred}{rgb}{0.9, 0.25, 0}
\definecolor{darkblue}{rgb}{0.188, 0.478, 0.9}

\let \savenumberline \numberline
\def \numberline#1{\savenumberline{#1.}}

\makeatletter
\def\@fpheader{\relax}
\makeatother

\def\bea{\begin{eqnarray}}
\def\eea{\end{eqnarray}}

\title{\ \vspace{1.6cm} \\
\scalebox{0.95}{KLT Factorization of Winding String Amplitudes}}
\author{Jaume Gomis${}^a$, Ziqi Yan${}^{b}$, and Matthew Yu${}^a$}
\affiliation{
${}^a$ Perimeter Institute for Theoretical Physics\\
        31 Caroline St N, Waterloo, ON N2L 2Y5, Canada \medskip\\
${}^b$ Nordita, KTH Royal Institute of Technology and Stockholm University\\
Hannes Alfv\'{e}ns v\"{a}g 12, SE-106 91 Stockholm, Sweden \medskip\\
E-mail: \emph{jgomis@pitp.ca, ziqi.yan@su.se, myu@pitp.ca}\\
\ \vspace{-0.8cm}}
\abstract{We uncover a Kawai-Lewellen-Tye (KLT)-type factorization of  closed string  amplitudes  into open string amplitudes for  closed string  states carrying winding and momentum in toroidal compactifications.  
The winding  and  momentum closed string  quantum numbers map respectively to the  integer   and fractional winding quantum numbers of open strings  ending on a D-brane  array localized in the compactified directions. The closed string amplitudes factorize  into products of open string scattering amplitudes with the open strings ending   on a  D-brane configuration determined   by closed string data.}

\begin{document}

\maketitle
\vfill\eject


\section{Introduction}

The celebrated Kawai-Lewellen-Tye (KLT) relations show  that   tree-level closed string amplitudes factorize  into a sum of quadratic products of open string amplitudes \cite{Kawai:1985xq}. Intuitively, the KLT relations are a consequence of the factorization of complex integrals describing closed string amplitudes into contour integrals describing open string amplitudes. The   simplest  incarnation of  such a factorization is the  classic  Riemann bilinear identity. 

The KLT relations have revealed elegant structures in scattering amplitudes and led to surprising connections between different theories (see e.g. \cite{BjerrumBohr:2009rd, Stieberger:2009hq, BjerrumBohr:2010hn, Cachazo:2013gna, Mafra:2016mcc, He:2016mzd, Mizera:2016jhj, Mizera:2017cqs, Mizera:2017rqa, Casali:2019ihm}).
Originally, the KLT relations were proposed as a powerful method for evaluating tree-level closed string amplitudes \cite{Kawai:1985xq}. 
Intriguingly, the discovery of such relations has also led to numerous insights into quantum field theory (QFT) amplitudes by considering the infinite tension limit of string amplitudes.
For example, it has been shown that Born-Infeld, non-linear sigma models, and special Galileons fit into the QFT type of KLT relations \cite{Bern:1998sv}, via the framework of the CHY formalism \cite{Cachazo:2014xea}.
The KLT relations also inspired studies of gravity as a double copy of gauge theory over the past decade  \cite{Bern:2010ue}. The importance of the double-copy construction is justified by its use in significantly simplifying the calculation of gravity amplitudes at loop level \cite{Bern:2008qj, Carrasco:2011mn}.
A deeper understanding of these remarkable relations may hold  new surprises in simplifying practical computations of amplitudes in both field and string theory. 

Connecting string theory with nature requires   compactifying string theory  down to four spacetime dimensions.  Upon compactification, the extended nature of the string     clearly  distinguishes string theory   from local QFT. The existence of string states winding along a compact direction, which do not have a local field theory description, enriches the quantum numbers carried by string states and the collection of string amplitudes  in the theory, and leads to   inherently stringy phenomena, such as T-duality. 

\begin{figure}[t!]
\centering
\begin{tikzpicture}
    \draw[line width = .3mm] (0,0) circle (1.5);
    \draw[decoration={markings, mark=at position .1 with {\arrow{>}}}, postaction={decorate}, line width=.3mm] (-4+1,1.5) -- (1,0); 
    \draw[decoration={markings, mark=at position .9 with {\arrow{<}}}, postaction={decorate}, line width=.3mm] (-1,0) -- (4-1,-1.5);
    \draw[decoration={markings, mark=at position .1 with {\arrow{>}}}, postaction={decorate},line width=.3mm] (-4.5+1,0.8) -- (1,0);
    \draw [ decoration={markings, mark=at position .9 with {\arrow{<}}}, postaction={decorate},line width=.3mm] (-1,0) -- (4.5-1,-0.8);
    \draw[decoration={markings, mark=at position .1 with {\arrow{>}}}, postaction={decorate}, line width=.3mm] (-3.65,0) -- (1,0); 
    \draw[decoration={markings, mark=at position .9 with {\arrow{<}}}, postaction={decorate},line width=.3mm] (-1,0) -- (3.65,0);
    \node at (-3,-0.3) {\scalebox{0.4}{$\bullet$}};
    \node at (-2.98,-0.5) {\scalebox{0.4}{$\bullet$}};
    \node at (-2.95,-0.7) {\scalebox{0.4}{$\bullet$}};
    \node at (-2.89,-0.9) {\scalebox{0.4}{$\bullet$}};
    \node at (-2.8,-1.1) {\scalebox{0.4}{$\bullet$}};
    \node at (3,0.3) {\scalebox{0.4}{$\bullet$}};
    \node at (2.98,0.5) {\scalebox{0.4}{$\bullet$}};
    \node at (2.95,0.7) {\scalebox{0.4}{$\bullet$}};
    \node at (2.89,0.9) {\scalebox{0.4}{$\bullet$}};
    \node at (2.8,1.1) {\scalebox{0.4}{$\bullet$}};
    \draw[decoration={markings, mark=at position .1 with {\arrow{>}}}, postaction={decorate},line width=.3mm] (-4+1,-1.5) -- (1,0);
    \draw[decoration={markings, mark=at position .9 with {\arrow{<}}}, postaction={decorate},line width=.3mm] (-1,0) -- (4-1,1.5);
    \draw[fill=blue!20,line width = .3mm] (0,0) circle (1.5);
    \draw (-3.7,1.5) node {\scalebox{.9}{$(n_1, w_1)$}};
    \draw (-4.2,0.8) node {\scalebox{.9}{$(n_2, w_2)$}};
    \draw (-4.4,0) node {\scalebox{.9}{$(n_3, w_3)$}};
    \draw (-3.7,-1.5) node {\scalebox{.9}{$(n_{k}, w_k)$}};
    \draw (3.85,1.5) node {\scalebox{.9}{$(n_{\CN}, w_{\CN})$}};
    \draw (4.6,-.8) node {\scalebox{.9}{$(n_{k+2}, w_{k+2})$}};
    \draw (4.8,0) node {\scalebox{.9}{$(n_{k+3}, w_{k+3})$}};
    \draw (4.1,-1.5) node {\scalebox{.9}{$(n_{k+1}, w_{k+1})$}};
    \draw (-6.5,0) node {\scalebox{0.9}{\emph{incoming}}};
    \draw (7,0) node {\scalebox{0.9}{\emph{outgoing}}};
\end{tikzpicture}
\vspace{-6mm}
$$
    n_+ = \sum{\phantom{\big|}\!\!}_{i=1}^k n_i
$$
\vspace{-9mm}
\caption{\small Closed string amplitude  with momentum   and winding  data  $(n_i,w_i)$ obeying      conservation laws $\sum_{i = 1}^\CN n_i = 0$ and $\sum_{i=1}^\CN w_i = 0$\,.  We  choose $n_1\, , \ldots, n_k> 0$ and $n_{k+1}\, , \ldots, n_{\CN} \leq 0$,      splitting strings into  incoming and  outgoing states. The integer $n_+$ is the  total incoming momentum.  We show  that   the number of D-branes   in  the  open string scattering amplitude appearing in the KLT-like factorization is $n_++1$.}
\label{fig:css}
\end{figure}
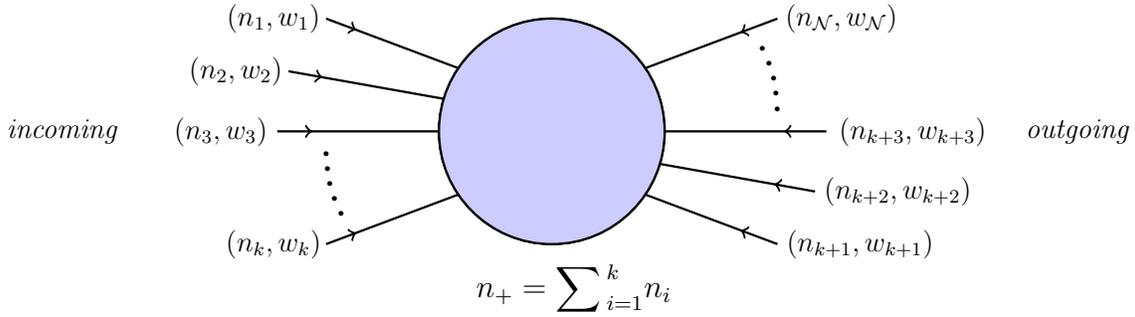
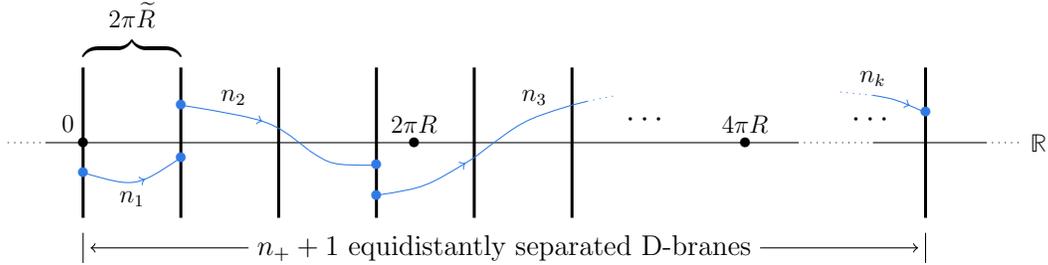
\begin{figure}[t!]
\centering
\begin{tikzpicture}
    \draw (-0.5,0)--(9.5,0) (10.5,0) -- (12,0);
    \draw (12.7,0) node {\scalebox{0.8}{$\mathbb{R}$}};
    \draw (0,0) node {\scalebox{.8}{$\bullet$}};
    \draw (-0.2,.25) node {\scalebox{.8}{$0$}};
    \draw (2*2.2,0) node {\scalebox{.8}{$\bullet$}};
    \draw (4*2.2,0) node {\scalebox{.8}{$\bullet$}};
    \draw (2*2.2,.25) node {\scalebox{.8}{$2\pi R$}};
    \draw (4*2.2,.25) node {\scalebox{.8}{$4\pi R$}};
    \draw (0.65,1.25) node {$\overbrace{\hspace{1.3cm}}$};
    \draw (0.65,1.7) node {\scalebox{0.8}{$2\pi \tilde R$}};
    \draw[line width = .4mm] (0,-1)--(0,1);
    \draw[line width = .4mm] (1.3,-1)--(1.3,1);
    \draw[line width = .4mm] (2*1.3,-1)--(2*1.3,1);
    \draw[line width = .4mm] (3*1.3,-1)--(3*1.3,1);
    \draw[line width = .4mm] (4*1.3,-1)--(4*1.3,1);
    \draw[line width = .4mm] (5*1.3,-1)--(5*1.3,1);
    \draw[line width = .4mm] (11.2,-1)--(11.2,1);
    \draw[line width= .4mm] (7.5,.3) node {$\dotsb$};
    \draw[line width= .4mm] (10.5,.3) node {$\dotsb$};
    \draw[dotted] (-1,0) -- (-0.5,0);
    \draw[dotted] (9.5,0) -- (10.5,0);
    \draw[dotted] (12,0) -- (12.5,0);
    \draw (0,-1.2) -- (0,-1.6);
    \draw (11.2,-1.2) -- (11.2,-1.6);
    \draw [->] (2.2,-1.4) -- (0.1,-1.4);
    \draw [->] (9,-1.4) -- (11.1,-1.4);
    \draw (5.6,-1.4) node {\scalebox{0.9}{$n_++1$ equidistantly separated D-branes}};
    \draw[darkblue] (1.3, 0.5) node {\scalebox{.8}{$\bullet$}};
    \draw[darkblue] (3*1.3, -0.3) node {\scalebox{.8}{$\bullet$}};
    \draw[darkblue] (0, -0.4) node {\scalebox{.8}{$\bullet$}};
    \draw[darkblue] (1.3, -0.2) node {\scalebox{.8}{$\bullet$}};
    \draw[darkblue] (3*1.3, -0.7) node {\scalebox{.8}{$\bullet$}};
    \draw[darkblue] (11.2, 0.4) node {\scalebox{.8}{$\bullet$}};
    \draw[darkblue,decoration={markings, mark=at position .6 with {\arrow{>}}}, postaction={decorate}] (0, -0.4).. controls (0.7,-0.6) .. (1.3,-0.2);
    \draw[darkblue,decoration={markings, mark=at position .6 with {\arrow{>}}}, postaction={decorate}] (1.3,0.5).. controls (2.5,0.3) .. (3,-0.1);
    \draw[darkblue] (3,-0.1).. controls (3.3,-0.3) .. (3*1.3,-0.3);
    \draw[darkblue,decoration={markings, mark=at position .9 with {\arrow{>}}}, postaction={decorate}] (3*1.3,-0.7).. controls (3.5*1.3,-0.6) .. (4*1.3,-0.2);
    \draw[darkblue] (4*1.3,-0.2).. controls (4.5*1.3,0.3) .. (5*1.3,0.5);
    \draw[darkblue] (5*1.3,0.5) -- (5*1.3+0.2,0.55);
    \draw[darkblue,dotted] (5*1.3+0.2,0.55) -- (5*1.3+0.6,0.63);
    \draw[darkblue,decoration={markings, mark=at position .4 with {\arrow{<}}}, postaction={decorate}] (11.2,0.4) .. controls (10.8,0.58) .. (10.5,0.63);
    \draw[darkblue,dotted] (10.5,0.63) -- (10,0.68);
    \draw (0.65, -0.75) node {\scalebox{0.8}{$n_1$}};
    \draw (2,.6) node{\scalebox{.8}{$n_2$}};
    \draw (6,.6) node{\scalebox{.8}{$n_3$}};
    \draw (10.5,.85) node{\scalebox{.8}{$n_k$}};
\end{tikzpicture}
\caption{\small The horizontal line represents the real axis, which under the  $x \sim x + 2 \pi R$ identification becomes a circle of radius $R$\,. The vertical lines represent an array of equidistantly separated D-branes. The distance between consecutive D-branes is $2\pi \tilde{R}$\,, where $\tilde{R} = \alpha' / R$ is the T-dual radius. Each blue curve  represents an incoming open string  stretched between different D-branes. The momentum $n_i$ of the $i$-th closed string  is encoded
in  the number of D-branes traversed by the $i$-th open string and physically corresponds to fractional winding.}
\label{fig:dbconfig}
\end{figure}

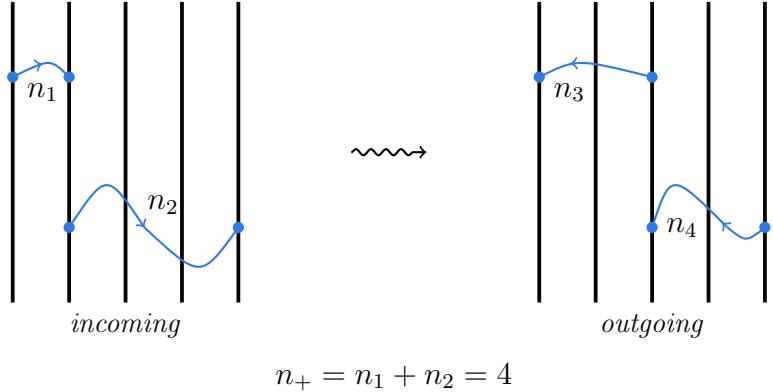
\begin{figure}[t!]
\centering
\begin{tikzpicture}[thick]
    \draw[black,line width= .5mm] (-.5-3,-2) -- (-.5-3,2) ;
    \draw[black,line width= .5mm] (0.25-3,-2) -- (0.25-3,2) ;
    \draw[black,line width= .5mm] (1-3,-2) -- (1-3,2) ;
    \draw[black,line width= .5mm] (1.75-3,-2) -- (1.75-3,2) ;
    \draw[black,line width= .5mm] (2.5-3,-2) -- (2.5-3,2) ;
    \tikzstyle{vertex}=[circle,fill=darkblue!100, minimum size=1.5pt,inner sep=1.5pt]
    \node[vertex]at ( -3.5 , 1) {};
    \node[vertex]at ( -2.75 , 1) {};
    \node[vertex]at ( -2.75 , -1) {};
    \node[vertex]at ( -.5 , -1) {};
    \draw[darkblue, decoration={markings, mark=at position 0.5 with
        {\arrow{>}}}, postaction={decorate}] (-3.5,1).. controls (-3,1.25) .. (-2.75,1);
    \draw[darkblue, decoration={markings, mark=at position 1 with
        {\arrow{>}}}, postaction={decorate}] (-2.75,-1).. controls (-2.25,-.25) .. (-1.75,-1);
    \draw[darkblue] (-1.75,-1).. controls (-1,-1.7) .. (-.5,-1);
    \draw[->,decorate,decoration={snake,amplitude=.4mm,segment length=2mm,post length=1mm}] (1,0) -- (2, 0);
    \draw[black,line width= .5mm] (-.5-3+7,-2) -- (-.5-3+7,2) ;
    \draw[black,line width= .5mm] (0.25-3+7,-2) -- (0.25-3+7,2) ;
    \draw[black,line width= .5mm] (1-3+7,-2) -- (1-3+7,2) ;
    \draw[black,line width= .5mm] (1.75-3+7,-2) -- (1.75-3+7,2) ;
    \draw[black,line width= .5mm] (2.5-3+7,-2) -- (2.5-3+7,2) ;
    \tikzstyle{vertex}=[darkblue, circle,fill=darkblue!100, minimum size=1.5pt,inner sep=1.5pt]
    \node[vertex]at (-3.5+7, 1) {};
    \node[vertex]at (-2+7, 1) {};
    \node[vertex]at (-2+7, -1) {};
    \node[vertex]at (-.5+7, -1) {};
    \draw[darkblue, decoration={markings, mark=at position 0.35 with
        {\arrow{<}}}, postaction={decorate}] (-3.5+7,1).. controls (-3+7,1.25) .. (-2+7,1);
    \draw[darkblue,decoration={markings, mark=at position 1 with
        {\arrow{<}}}, postaction={decorate}] (-2+7,-1).. controls (-1.75+7,-.25) .. (-1+7,-1);
    \draw[darkblue] (-1+7,-1).. controls (-.75+7,-1.2) .. (-.5+7,-1);
    \node at (-3.1,0.8) {$n_1$};
    \node at (-3.1+7,0.8) {$n_3$};
    \node at (-1.5,-0.7) {$n_2$};
    \node at (-1.5+6.9,-1) {$n_4$};
    \node at (-2,-2.3) {\scalebox{0.9}{\emph{incoming}}};
    \node at (5,-2.3) {\scalebox{0.9}{\emph{outgoing}}};
\end{tikzpicture}
\vspace{-5mm}
$$
    n_+ = n_1 + n_2 = 4
$$
\vspace{-1cm}
\caption{\small A four-point open string amplitude with $n_1 = 1$\,, $n_2 = 3$\,, $n_3 = -2$\,, and $n_4 = -2$\,. Open strings can only join or split at the same D-brane. There is a total of $n_+ + 1 = 5$ D-branes for this choice of closed string quantum numbers.}
\label{fig:osamp}
\end{figure}

In this paper we derive a  KLT-like relation for $\mathcal{N}$-point scattering amplitudes of closed strings when a spatial direction is compactified on a circle and string states carry winding (and momentum) on the circle. This allows us to express any closed string amplitude in terms of open string amplitudes. But which open string amplitudes? At first sight there seems to be a puzzle. A closed string state  on a circle of radius $R$ can carry both momentum $n / R$ and winding $w$ along that circle, where $n, \, w \in \mathbb{Z}$\,. In contrast, an open string state can either carry momentum or winding along the circle, depending upon whether we impose Neumann or Dirichlet boundary conditions along the circle, but {\it cannot} carry both momentum and winding. The resolution of this conundrum is that, while the winding number $w_i$ of the $i$-th closed string can be identified with the winding number of the $i$-th open string  obeying Dirichlet boundary conditions, the closed string momentum $ n_i$ is encoded in the D-brane configuration where open string amplitudes are defined, and corresponds to the fractional winding number of an open string stretched between D-branes. Figure \ref{fig:css} shows the scattering of $\CN$ closed strings with momentum and winding data $(n_i,w_i)$\,. Figure \ref{fig:dbconfig} illustrates the corresponding D-brane configuration. The D-branes are equally separated by a distance which is   T-dual to the circumference of the compactified circle. The associated open string amplitude  is   depicted in Figure \ref{fig:osamp}, and  involves open strings  defined on an array of $n_++1$ D-branes, with $n_+$ defined in Figure \ref{fig:css}. The $i$-th open string winds $w_i$ times around the full spatial circle; in addition, it traverses $n_i - 1$ D-branes, with $n_i$ corresponding to the ``fractional" part of winding number of the open string stretched between two D-branes. The fractional winding number is equal to $n_i \, \tilde R/R=\alpha' \, n_i / R^2$\,. The above mapping between the quantum numbers describing closed and open string states  in the KLT-like relation is summarized in the table below:
$$
\begin{array}{cc}
   \text{\bf{closed string}}\quad \phantom{\longleftrightarrow} & \text{\bf{open string}}  \\[1pt] \hline\\[-11pt] 
   \text{momentum}  \quad \phantom{\longleftrightarrow} & \text{fractional winding}\\
   \text{\phantom{w}\,\,winding}  \qquad \phantom{\longleftrightarrow} & \text{integer winding} 
\end{array}
$$
The precise meaning of these terms will be further clarified in \S\ref{sec:KLTfirst}.\,\footnote{There is also a straightforward T-dual interpretation for open string scattering amplitudes with open strings ending on the D-brane configuration described in Figure \ref{fig:dbconfig}. In the T-dual frame, an open string ends on a stack of spacetime-filling D-branes   in the presence of Wilson lines. Such an open string satisfies Neunmann boundary conditions in the compactified direction and can carry nonzero momentum number (both integral and fractional, with the latter attributed to the Wilson line) but only zero winding. See \S\ref{sec:fpKLT} for details.} 

This paper is organized as follows. In \S\ref{sec:ewcs}, we give an introduction to essential ingredients of winding closed strings, and evaluate in \S\ref{sec:4ptcs} the four closed string tachyon~\footnote{Although a closed string tachyon cannot simultaneously have nonzero momentum and winding along the compactified circle, the four-tachyon amplitude still provides a sufficiently interesting example that captures the salient properties of the KLT relation for winding strings.} amplitude in the presence of windings. In \S\ref{sec:osKLT}, we discuss winding open strings and derive in \S\ref{sec:fpKLT} the KLT relation between four closed and open string tachyon amplitudes. In \S\ref{sec:hpa}, we generalize the KLT relation for winding strings to higher-point amplitudes that involve  massless and massive vertex operators, and present the general form of the KLT relation in \eqref{eq:Mcresult}. We conclude our paper in \S\ref{sec:conclusions}. In Appendix \ref{app:Rpr}, we recast our new KLT relation in more modern terms using intersection theory, which gives rise to a compact rewriting of the KLT relation in \eqref{eq:itKLT}.  

\section{Elements on Winding Closed Strings} \label{sec:ewcs}

We consider string theory described by a sigma model that maps the worldsheet $\Sigma$ to a $d$-dimensional target space $\CM$\,. We take the worldsheet to be Euclidean and denote the worldsheet coordinates by $\sigma^\alpha = (\tau, \sigma)$\,, with $\tau \in \mathbb{R}$ and $\sigma \in [0,2\pi]$\,. In the target space, we denote the coordinates by $\mathbb{X}^M = (X^\mu, X)$\,, where $M = 0, \ldots, d-1$ and $\mu = 0\,, 1\,, \ldots, d-2$\,. We have defined $X^\mu \equiv \mathbb{X}^\mu$ and $X \equiv \mathbb{X}^{d-1}$. We compactify $X$ over a circle of radius $R$\,, such that
\be
	X \sim X + 2 \pi R\,.
\ee
A closed string that winds $w$ times around the circle along the $X$-direction satisfies
\be
	X (\sigma +2 \pi) = X(\sigma) + 2 \pi R \, w\,,
		\qquad
	w \in \mathbb{Z}\,.
\ee
The mode expansion of $X$ is
\be
	X (\tau, \sigma) = x - i \alpha' p \, \tau + R \, w \, \sigma + i \sqrt{\frac{\alpha'}{2}} \sum_{m\neq 0} \frac{1}{m} \lr \alpha_m \, e^{- i m \sigma} + \tilde{\alpha}_m \, e^{i m \sigma} \rr e^{- m \tau},
\ee
where $p$ denotes the momentum in the $X$-direction and $\alpha'$ is the Regge slope. The single-valuedness of the operator $\exp (i p X)$ requires the quantization condition
\be
	p = \frac{n}{R}\,,
		\qquad
	n \in \mathbb{Z}\,.
\ee
Here, $n$ is the Kaluza-Klein (KK) excitation number. 

To perform radial quantization, we define
\be \label{eq:zzdef}
	z = e^{\tau + i \sigma}\,,
		\qquad
	\overline{z} = e^{\tau - i \sigma}\,,
\ee 
in terms of which the mode expansion of $X$ is $X (z\,, \overline{z}) = X_{\text{L}} (z) + X_{\text{R}} (\overline{z})$\,,
where
\begin{subequations}
\begin{align}
	X_{\text{L}} (z) & = x_{\text{L}} - \frac{i}{2} \, \alpha' p_{\text{L}} \ln z + i \sqrt{\frac{\alpha'}{2}} \sum_{m\neq0} \frac{\alpha_m}{m \, z^m}\,, \\[2pt]
	X_{\text{R}} (\overline{z}) & = x_{\text{R}} - \frac{i}{2} \, \alpha' p_{\text{R}} \ln \overline{z} + i \sqrt{\frac{\alpha'}{2}} \sum_{m\neq0} \frac{\tilde{\alpha}_m}{m \, {\overline{z}}^m}\,, 
\end{align}
\end{subequations}
with $x = x_\text{L} + x_\text{R}$ and
\be
	p^{}_{\text{L}} = \frac{n}{R} + \frac{R \, w}{\alpha'}\,,
		\qquad
	p^{}_{\text{R}} = \frac{n}{R} - \frac{R \, w}{\alpha'}\,.
\ee
Note that $p_{\text{L}}$ and $p_{\text{R}}$ are eigenvalues of the operators
\be
	\hat{p}^{}_{\text{L}} = \frac{1}{\pi \alpha'} \oint_\CC dz  \, \p_z X_{\text{L}} (z)\,,
		\qquad
	\hat{p}^{}_{\text{R}} = - \frac{1}{\pi \alpha'} \oint_\CC d\overline{z} \, \p_{\overline{z}} X_{\text{R}} (\overline{z})\,,
\ee
where the counterclockwise oriented contour $\CC$ encloses the vertex operator  associated with the string. 

\subsection{Closed String Tachyons and Cocycles}

The closed string tachyon is described by the following vertex operator \cite{Polchinski:1998rq}:
\begin{align} \label{eq:cvo}
\begin{split}
	\CV_\CC (z\,, \overline{z}) & = g_\text{c} \exp \bigl[ \tfrac{i}{4} \, \pi \alpha' (p^{}_{\text{L}} - p^{}_{\text{R}}) (\hat{p}^{}_{\text{L}} + \hat{p}^{}_{\text{R}}) \bigr] : \! e^{i K_{\text{L}} \cdot \mathbb{X}_{\text{L}} (z) + i K_{\text{R}} \cdot \mathbb{X}_{\text{R}} (\overline{z})} \! : \\[4pt]
	& = g_\text{c} \exp \bigl[ \tfrac{i}{2} \, \pi R \, w \, (\hat{p}^{}_{\text{L}} + \hat{p}^{}_{\text{R}}) \bigr] : \! e^{i K_{\text{L}} \cdot \mathbb{X}_{\text{L}} (z) + i K_{\text{R}} \cdot \mathbb{X}_{\text{R}} (\overline{z})} \! :\,,
\end{split}
\end{align}
where $K^{}_{\text{L}, \, \text{R}} \cdot \mathbb{X}^{}_{\text{L}, \, \text{R}} = \eta^{}_{MN} \, K^M_{\text{L}, \, \text{R}} \, \mathbb{X}^N_{\text{L}, \, \text{R}}$\,, and
\begin{subequations} 
\begin{align}
	\mathbb{X}_{\text{L}}^M & = \lr X^\mu_{\phantom{L}}, \, X^{}_{\text{L}} \rr,
		&
	K_\text{L}^M & = \lr k^\mu_{\phantom{L}}, \, p^{}_{\text{L}} \rr, \label{eq:KLMdef} \\[2pt]
	\mathbb{X}_{\text{R}}^M & = \lr X^\mu_{\phantom{L}}, \, X^{}_{\text{R}} \rr,
		&
	K_\text{R}^M & = \lr k^\mu_{\phantom{L}}, \, p^{}_{\text{R}} \rr. \label{eq:KRMdef}
\end{align}
\end{subequations}
The extra phase factor in \eqref{eq:cvo} is known as the \emph{cocycle}, which is important for removing the phases arising from crossing certain branch cuts when different vertex operators are interchanged with each other.

To demonstrate that \eqref{eq:cvo} is the correct vertex operator, we first note the operator product expansions (OPEs),
\begin{subequations}
\begin{align}
	X_{\text{L}} (z_1) \, X_{\text{L}} (z_2) & \sim - \frac{\alpha'}{2} \, \ln z_{12}\,, 
		&
	X_{\text{L}} (z_1) \, X^\mu (z_2\,, \overline{z}_2) & \sim 0\,, \\[2pt]
	X_{\text{R}} (\overline{z}_1) \, X_{\text{R}} (\overline{z}_2) & \sim - \frac{\alpha'}{2} \, \ln \overline{z}_{12}\,, 
		&
	X_{\text{R}} (\overline{z}_1) \, X^\mu (z_2\,, \overline{z}_2) & \sim 0\,, \\[8pt]
	X_{\text{L}} (z_1) \, X_{\text{R}} (\overline{z}_2) & \sim 0\,, 
\end{align}
and
\be
	X^\mu (z_1\,, \overline{z}_1) \, X^\nu (z_2\,, \overline{z}_2) \sim - \frac{\alpha'}{2} \, \eta^{\mu\nu} \, \ln |z_{12}|^2.
\ee
\end{subequations}
We defined $z_{12} \equiv z_1 - z_2$ and $\overline{z}_{12} = \overline{z}_1 - \overline{z}_2$\,.
It follows that
\be
	\hat{p}^{}_{\text{L}} \CV_\CC (z\,, \overline{z}) = \frac{1}{\pi \alpha'} \oint_{\mathcal{C}} dw : \! \p_w X_{\text{L}} (w) \! : \CV_\CC (z\,, \overline{z}) = p^{}_{\text{L}} \CV_\CC (z\,, \overline{z})\,.
\ee
Therefore, $\CV_\CC (z\,, \overline{z})$ corresponds to the eigenstate of $\hat{p}^{}_{\text{L},\,\text{R}}$\,. Similarly, we have
\be
	\hat{p}^{}_{\text{R}} \CV_{\CC}(z\,, \overline{z}) = p^{}_{\text{R}} \CV_{\CC}(z\,, \overline{z})\,.
\ee

We next note that the BRST invariance of \eqref{eq:cvo} requires
\be
	- K_\text{L}^2 = - K_\text{R}^2 = - \frac{4}{\alpha'}
		\qquad \implies \qquad
	- k^2 = p_{\text{L}}^2 - \frac{4}{\alpha'} = p_{\text{R}}^2 - \frac{4}{\alpha'}\,,  
\ee
where $k^2 = k_\mu k^\mu$\,. It then follows that the dispersion relation is
\be
	- k^2 = \frac{n^2}{R^2} + \frac{w^2 R^2}{{\alpha'}^2} - \frac{4}{\alpha'}\,,
\ee
when supplemented with the level matching condition 
\be \label{eq:nw0}
	n \, w = 0\,.
\ee
This level matching condition implies that an asymptotic tachyonic state cannot have both nonzero Kaluza-Klein and winding numbers.

Finally, we consider the product of two tachyon operators, such as
\be
	\CV_{\CC_1} (z_1\,, \overline{z}_1) \, \CV_{\CC_2} (z_2\,, \overline{z}_2)\,,
\ee
which is taken to be radially ordered, i.e. $\CC_1$ encloses $\CC_2$\,. It follows that
\begin{align}
\begin{split}
	& \quad \CV_{\CC_1} (z_1\,, \overline{z}_1) \, \CV_{\CC_2} (z_2\,, \overline{z}_2) \\[2pt]
	& \sim g_\text{c}^2 \, e^{i \pi \, w_1 n_2} \, z_{12}^{\frac{1}{2} \alpha' K_{\text{L}1} \cdot K_{\text{L}2}} \, \, \overline{z}_{12}^{\frac{1}{2} \alpha' K_{\text{R}1} \cdot K_{\text{R}2}} \, : \! e^{i(K_{\text{L}1} + K_{\text{L}2}) \cdot \mathbb{X}_{\text{L}} (z) + i(K_{\text{R}1} + K_{\text{R}2}) \cdot \mathbb{X}_{\text{R}} (\overline{z})} \! : \\[2pt]
	& = g_\text{c}^2 \, e^{i \pi \, w_1 n_2} \, \bigl( e^{\frac{1}{2} i \pi \alpha' K_{\text{L}1} \cdot K_{\text{L}2}} \, z_{21}^{\frac{1}{2} \alpha' K_{\text{L}2} \cdot K_{\text{L}1}} \bigr) \, \bigl( e^{- \frac{1}{2} i \pi \alpha' K_{\text{R}1} \cdot K_{\text{R}2}} \, \overline{z}_{21}^{\frac{1}{2} \alpha' K_{\text{R}2} \cdot K_{\text{R}1}} \bigr) \\[2pt]
	& \hspace{6.2cm} \times : \! e^{i(K_{\text{L}2} + K_{\text{L}1}) \cdot \mathbb{X}_{\text{L}} (z) + i(K_{\text{R}2} + K_{\text{R}1}) \cdot \mathbb{X}_{\text{R}} (\overline{z})} \! :\,.
\end{split}
\end{align}
Using
\be \label{eq:KL12-KR12}
	\tfrac{1}{2} \, \alpha' \bigl( K_{\text{L}1} \cdot K_{\text{L}2} - K_{\text{R}1} \cdot K_{\text{R}2} \bigr) = n_1 \, w_2 + n_2 \, w_1\,,
\ee
we find
\begin{align}
	& \quad \CV_{\CC_1} (z_1\,, \overline{z}_1) \, \CV_{\CC_2} (z_2\,, \overline{z}_2) \notag \\[2pt]
	& \sim e^{2 \, i \pi \, w_1 n_2} \Bigl[ g_\text{c}^2 \, e^{i \pi \, w_2 \, n_1} z_{21}^{\frac{1}{2} \alpha' K_{\text{L}2} \cdot K_{\text{L}1}} \,\, \overline{z}_{21}^{\frac{1}{2} \alpha' K_{\text{R}2} \cdot K_{\text{R}1}} \, : \! e^{i(K_{\text{L}2} + K_{\text{L}1}) \cdot \mathbb{X}_{\text{L}} (z) + i(K_{\text{R}2} + K_{\text{R}1}) \cdot \mathbb{X}_{\text{R}} (\overline{z})} \! : \Bigr] \notag \\[2pt]
	& \sim e^{2 \, i \pi \, w_1 n_2} \, \CV_{\CC_2} (z_2\,, \overline{z}_2) \, \CV_{\CC_1} (z_1\,, \overline{z}_1)\,.
\end{align}
Since both $w_1$ and $n_2$ are integers, we find that $\CV_{\CC_1} (z_1\,, \overline{z}_1)$ and $\CV_{\CC_2} (z_2\,, \overline{z}_2)$ commute. 

\subsection{Higher-Order Vertex Operators} \label{sec:hovo}

More general vertex operators can be constructed by multiplying vertex operator with  factors of the form $\p^m \mathbb{X}^M_{\text{L}}$ and $\bar{\p}{}^m \mathbb{X}^M_{\text{R}}$, for positive integer $m$\,.  BRST invariance imposes various conditions on such operators. For example, we have the BRST-invariant (up to total derivatives) and gauge-fixed vertex operators
\begin{align} \label{eq:gvo1}
	& e^{\tfrac{1}{2} \, i \pi R \, w \, (\hat{p}^{}_{\text{L}} + \hat{p}^{}_{\text{R}})}  \, \varepsilon^{}_{I_1 \cdots I_N \, J_1 \cdots J_{\tilde{N}}} \notag \\[4pt]
		& \qquad \times : \! \p \mathbb{X}_{\text{L}}^{I_1} \cdots \p \mathbb{X}_{\text{L}}^{I_N} \, \bar{\p} \mathbb{X}_{\text{R}}^{J_1} \cdots \bar{\p} \mathbb{X}_{\text{R}}^{J_{\tilde{N}}} \, e^{i K_{\text{L}} \cdot \mathbb{X}_{\text{L}} (z) + i K_{\text{R}} \cdot \mathbb{X}_{\text{R}} (\overline{z})} \! :\,,
\end{align}
where $\varepsilon^{}_{I_1 \cdots I_N \, J_1 \cdots J_{\tilde{N}}}$ is symmetric in $\{ I_1, \ldots, I_N \}$ and in $\{ I_1, \ldots, I_{\tilde{N}} \}$\,, respectively, and
\begin{subequations}
\begin{align}
	K_{\text{L}}^{I_i} \, \varepsilon^{}_{I_1 \cdots I_i \cdots I_N \, J_1 \cdots J_{\tilde{N}}} & = 0\,, 
		\qquad
	\eta^{I_i J_j} \, \varepsilon^{}_{I_1 \cdots I_i \cdots  I_N \, J_1 \cdots J_j \cdots J_{\tilde{N}}} = 0\,, \\[2pt]
	K_{\text{R}}^{J_j} \, \varepsilon^{}_{I_1 \cdots I_N \, J_1 \cdots J_j \cdots J_{\tilde{N}}} & = 0\,.
\end{align}
\end{subequations}
The dispersion relation and the level matching conditions are
\be \label{eq:gdrlm}
	- k^2 = \frac{n^2}{R^2} + \frac{w^2 R^2}{{\alpha'}^2} + \frac{2}{\alpha'} \, \bigl( N + \tilde{N} - 2 \bigr)\,,
		\qquad
	n \, w = \tilde{N} - N\,.
\ee
Unlike closed string tachyons, the vertex operator \eqref{eq:gvo1} can simultaneously have nonzero KK and winding quantum numbers.

For an example of a BRST-invariant vertex operator that contains higher derivatives acting on $X_\text{L,\,R}$\,, we consider the gauge-fixed operator
\begin{align} \label{eq:gvo2}
	& e^{\tfrac{1}{2} \, i \pi R \, w \, (\hat{p}^{}_{\text{L}} + \hat{p}^{}_{\text{R}})}  \, \varepsilon^{}_{M N} \, : \! \bigl( \tfrac{i}{2} \, \alpha' K^M_\text{L} \p^2 \mathbb{X}_\text{L}^N + \p \mathbb{X}^M_{\text{L}} \, \p \mathbb{X}^N_\text{L} \bigr) \, e^{i K_{\text{L}} \cdot \mathbb{X}_{\text{L}} (z) + i K_{\text{R}} \cdot \mathbb{X}_{\text{R}} (\overline{z})} \! :\,,
\end{align}
where $\varepsilon_{MN}$ is a symmetric tensor that satisfies
$\varepsilon^M{}_M = \alpha' K_{\text{L}}^M \, K_{\text{L}}^N \, \varepsilon^{}_{MN}$\,.
The dispersion relation and the level matching condition in this case are
\be
	- k^2 = \frac{n^2}{R^2} + \frac{w^2 R^2}{{\alpha'}^2}\,,
		\qquad
	n \, w = - 2\,.
\ee
Other higher-derivative vertex operators can be constructed similarly.

\subsection{Four Closed String Tachyon Amplitudes} \label{sec:4ptcs}

We start with the tree-level four tachyon amplitude, which exhibits the simplest nontrivial KLT relation. Due to the level matching condition \eqref{eq:nw0}, a tachyonic state can either have a nonzero KK or winding number. Later in \S\ref{sec:hpa}, we will consider amplitudes that involve vertex operators discussed in \S\ref{sec:hovo}, in which case an asymptotic closed string state can have both nonzero KK and winding numbers. The amplitude for four closed string tachyons on a spherical worldsheet is 
\begin{align} \label{eq:A4}
	\CA^{(4)}_\text{c} & = e^{- \chi \, \Phi_0} \, \int_\mathbb{C} d^2 z_2 \left \langle : \! c(z_1) \, \tilde{c} (\overline{z}_1) \! : \,\, : \! c(z_3) \, \tilde{c} (\overline{z}_3) \! : \,\, : \! c(z_4) \, \tilde{c} (\overline{z}_4) \! : \prod_{i=1}^4 : \! \CV_{\CC_i} (z_i\,, \overline{z}_i) \! : \right \rangle_{\!\!\text{S}^2} \notag \\[2pt]
	& \propto i \, (2\pi)^{25} \, \delta^{(25)} \bigl( k_1 + \cdots + k_4 \bigr) \, \delta_{n_1 + \cdots + n_4, \, 0} \, \delta_{w_1 + \cdots + w_4, \, 0} \, \CM^{(4)} \,.
\end{align}
Here, $\chi = 2$ is the Euler characteristic of the sphere and $\Phi_0$ is the expectation value of the dilaton field. 
We have dropped a proportionality constant that depends on the coupling $g_\text{c}$\,, which we do not need in this paper. The vertex operators $\CV_{\CC_i}$ are ordered radially, with the integration contour $\CC_i$ enclosing $\CC_j$\,, $i < j$\,. It follows that
\begin{align} \label{eq:M4}
	\CM^{(4)} & =  \frac{1}{\alpha'} \, |z_{13}|^2 \, |z_{14}|^2 \, |z_{34}|^2 \int_\mathbb{C} d^2 z_2 \prod_{\substack{i,\, j = 1 \\[2pt] i < j}}^4 e^{i \pi w_i n_j}  \, z^{\frac{1}{2} \alpha' K_{Li} \cdot K_{Lj}}_{ij} \, \overline{z}^{\frac{1}{2} \alpha' K_{\text{R}i} \cdot K_{\text{R}j}}_{ij}\,.
\end{align}
The vertex operator $\CV_{\CC_i} (z_i\,, \overline{z}_i)$ is defined in \eqref{eq:cvo}. We have used the M\"{o}bius transformations to fix 
$z_1 = 0$\,, $z_3 = 1$\,, and $z_4 = \infty$\,. Using the identity
\begin{align}
	z_{ij}^{\frac{1}{2} \alpha' K_{\text{L}i} \cdot K_{\text{L}j}} \, \overline{z}_{ij}^{\frac{1}{2} \alpha' K_{\text{R}i} \cdot K_\text{R}j} & = \bigl( e^{\frac{1}{2} i\pi \alpha' K_{\text{L}i} \cdot K_{\text{L}j}} \, z_{ji}^{\frac{1}{2} \alpha' K_{\text{L}i} \cdot K_{\text{Lj}}} \bigr) \, \bigl( e^{- \frac{1}{2} i\pi \alpha' K_{\text{R}i} \cdot K_{\text{R}j}} \, \overline{z}_{ji}^{\frac{1}{2} \alpha' K_{\text{R}i} \cdot K_{\text{Rj}}} \bigr) \notag \\[2pt]
		& = e^{i \pi (n_i w_j + n_j w_i)} \, z_{ji}^{\frac{1}{2} \alpha' K_{\text{L}i} \cdot K_{\text{Lj}}} \, \overline{z}_{ji}^{\frac{1}{2} \alpha' K_{\text{R}i} \cdot K_{\text{Rj}}},
\end{align}
we find that \eqref{eq:A4} becomes
\begin{align} \label{eq:Mc4I4}
	\CM_\text{c}^{(4)} 		
	& = \frac{1}{\alpha'} \, \exp \Bigl( i \pi {\sum}_{\substack{i,j=1 \\ i < j}}^4 \, n_i \, w_j \Bigr) \, \CI^{(4)} \,,
\end{align}
where
\be \label{eq:I4original}
	\CI^{(4)} = \int_{\mathbb{C}} d^2 z_2 \,\, z_2^{\frac{1}{2} \alpha' K_{\text{L}1} \cdot K_{\text{L}2}} \,\, \overline{z}_2^{\frac{1}{2} \alpha' K_{\text{R}1} \cdot K_{\text{R}2}} 
	\, (1- z_2)^{\frac{1}{2} \alpha' K_{\text{L}2} \cdot K_{\text{L}3}} \, (1 - \overline{z}_2)^{\frac{1}{2} \alpha' K_{\text{R}2} \cdot K_{\text{R}3}}{}.
\ee
This integral can be studied using intersection theory, which gives rise to a geometrical interpretation of the associated KLT relation in terms of the twisted (co)homology and intersection number. We will discuss this modern interpretation later in Appendix \ref{app:Rpr}. In practice, however, we will perform the integrals following the method used in the original work in \cite{Kawai:1985xq}, without resorting to intersection theory. This allows us to focus on the novelties brought by nonzero windings.

Define $z_2 = x + i \, \tilde{y}$ (and $\overline{z}_2 = x - i \, \tilde{y}$) for real variables $x$ and $\tilde{y}$\,, and then promote the integral over $\tilde{y}$ to be in the complex plane. Then, the integrand for $\tilde{y}$ has its branch points at $\pm i x$ and $\pm i (1 - x)$\,. We deform the contour for the $\tilde{y}$\,-integral that is along the real axis counterclockwise by $\frac{\pi}{2} - \epsilon$\,, for $\epsilon \rightarrow 0^+$. This deformation is captured by the following change of variables:
\be
	\tilde{y} = \exp \bigl[i ( \tfrac{\pi}{2} - \epsilon \bigr) \bigr] \, y \approx  i \, y + \epsilon \, y\,.
\ee
Therefore,\,\footnote{We use the convention $d^2 z_2 = dx \, d\tilde{y}$ in \eqref{eq:I4original}.}
\begin{align}
\begin{split}
	\CI^{(4)} & = \frac{i}{2} \int_\mathbb{R} d\zeta \int_\mathbb{R} d\xi \, ( \zeta + i \, \delta )^{\frac{1}{2} \alpha' K_{\text{L}1} \cdot K_{\text{L}2}} \, (1 - \zeta - i \, \delta)^{\frac{1}{2} \alpha' K_{\text{L}2} \cdot K_{\text{L}3}} \\[2pt]
		& \hspace{5cm} \times  ( \xi - i \, \delta )^{\frac{1}{2} \alpha' K_{\text{R}1} \cdot K_{\text{R}2}} \, ( 1 - \xi + i \, \delta )^{\frac{1}{2} \alpha' K_{\text{R}2} \cdot K_{\text{R}3}},
\end{split}
\end{align}
where 
\be
	\xi \equiv x + y\,, 
		\qquad
	\zeta \equiv x - y\,, 
		\qquad
	\delta \equiv \epsilon \, y = \tfrac{1}{2} \, \epsilon \, (\xi - \zeta)\,.
\ee
For a fixed $\zeta \in (- \infty\,, 0 )$\,, the $\xi$\,-integral has branch points at 
\be \label{eq:bp}
	i \, \delta \big|_{\xi = 0} = \tfrac{1}{2} \, i \, \epsilon \, ( - \zeta )\,,
		\qquad
	1 + i \, \delta \big|_{\xi = 1} = 1 + \tfrac{1}{2} \, i \, \epsilon \, (1 - \zeta)\,.
\ee
Importantly, both of the branch points reside in the upper half plane. Completing the integration contour for $\xi$ in the lower half plane, the integral over $\xi$ yields zero. A similar argument also shows that $\zeta \in (1, \infty)$ contributes zero to the amplitude. Therefore, only when $\zeta \in (0, 1)$\,, the amplitude receive nonzero contributions and
\begin{align} \label{eq:coo}
	\CI^{(4)} & = \frac{1}{2} \, i \, \CI_{\text{L}}^{(4)} \, \CI_\text{R}^{(4)}\,,
\end{align} 
where
\begin{subequations} \label{eq:ILR}
\begin{align}
	\CI_\text{L}^{(4)} & = \int_0^1 d\zeta \, \zeta^{\frac{1}{2} \alpha' K_{\text{L}1} \cdot K_{\text{L}2}} \, (1 - \zeta)^{\frac{1}{2} \alpha' K_{\text{L}2} \cdot K_{\text{L}3}}, \label{eq:ILR1} \\[2pt]
	\CI_\text{R}^{(4)} & = \int_{\mathbb{R}} d\xi \, (\xi - i \, \delta)^{\frac{1}{2} \alpha' K_{\text{R}1} \cdot K_{\text{R}2}} \, (1 - \xi + i \, \delta)^{\frac{1}{2} \alpha' K_{\text{R}2} \cdot K_{\text{R}3}}.
\end{align}
\end{subequations}
In $\CI_\text{R}^{(4)}$\,, there are two branch points at the positions given in \eqref{eq:bp}, but now with $\zeta \in (0, 1)$\,. This implies that the branch point at $i \, \delta \big|_{\xi = 0}$ resides in the lower half plane, and the branch point at $1 + i \, \delta \big|_{\xi = 1}$ resides in the upper half plane. 
We then take the part of the $\xi$\,-integral contour that lies to the right of $\xi = 0$\,, and rotate it by $180^\circ$ across the lower half plane of the complex plane, around the branch point $\xi = i \, \delta \big|_{\xi = 0}$\,. Note that this contour deformation does not cross any branch points. See \cite{Kawai:1985xq} for more details. We find
\begin{align} \label{eq:IR1}
	\CI_\text{R}^{(4)} & = 2 \, i \sin \bigl( \tfrac{1}{2} \, \pi \alpha' K_{\text{R}1} \cdot K_{\text{R}2} \bigr) \, \int_{-\infty}^0 d\xi \, (-\xi)^{\frac{1}{2} \alpha' K_{\text{R}1} \cdot K_{\text{R}2}} \, (1 - \xi)^{\frac{1}{2} \alpha' K_{\text{R}2} \cdot K_{\text{R}3}}\,.
\end{align}
We can alternatively take the part of the original $\xi$\,-integral contour that lies to the left of $\xi = 1$ to the right, and rotate is by $180^\circ$ across the upper half plane, around the branch point $1 + i \, \delta \big|_{\xi = 1}$\,. This contour deformation does not cross any branch points, either. This second choice of integration contour gives
\be \label{eq:IR2}
	\CI_{\text{R}}^{(4)} = 2 \, i \sin \bigl( \tfrac{1}{2} \, \pi \alpha' K_{\text{R}2} \cdot K_{\text{R}3} \bigr) \int_1^\infty d\xi \, \xi^{\frac{1}{2} \alpha' K_{\text{R}1} \cdot K_{\text{R}2}} \, (\xi - 1)^{\frac{1}{2} \alpha' K_{\text{R}2} \cdot K_{\text{R}3}}.
\ee

To proceed further, we introduce the generalized Mandelstam variables,
\begin{subequations}
\begin{align}
	s^{}_\text{L} & = - ( K_{\text{L}1} + K_{\text{L}2} )^2\,,
		&%
	t^{}_\text{L} & = - ( K_{\text{L}1} + K_{\text{L}3} )^2\,,
		&%
	u^{}_\text{L} & = - ( K_{\text{L}1} + K_{\text{L}4} )^2\,, \\[2pt]
	s^{}_\text{R} & = - ( K_{\text{R}1} + K_{\text{R}2} )^2\,,
		&%
	t^{}_\text{R} & = - ( K_{\text{R}1} + K_{\text{R}3} )^2\,,
		&%
	u_\text{R} & = - ( K_{\text{R}1} + K_{\text{R}4} )^2\,,
\end{align}
\end{subequations}
and define $\alpha_x \equiv - \frac{1}{4} \, \alpha' \, x - 1$\,. Note that $\alpha_{s^{}_\text{L}} + \alpha_{t^{}_\text{L}} + \alpha_{u^{}_\text{L}} = 1$ and $\alpha_{s^{}_\text{R}} + \alpha_{t^{}_\text{R}} + \alpha_{u^{}_\text{R}} = 1$\,. It follows that
\begin{subequations}
\begin{align}
	\CI_\text{L}^{(4)} & = B (\alpha_{s^{}_\text{L}}\,, \alpha_{u^{}_\text{L}} )\,, \\[2pt]
	\CI_\text{R}^{(4)} & = - 2 \, i \sin (\pi \, \alpha_{s^{}_\text{R}} ) \, B ( \alpha_{s^{}_\text{R}}, \alpha_{t^{}_\text{R}} ) \notag \\[2pt]
	& = - 2 \, i \sin ( \pi \, \alpha_{u^{}_{\text{R}}} ) \, B ( \alpha_{u^{}_{\text{R}}}, \alpha_{t^{}_\text{R}} )\,, \label{eq:IR4}
\end{align}
\end{subequations}
where $B (a, b)$ is the Euler beta function, with
\be
	B(a, b) = \frac{\Gamma(a) \, \Gamma(b)}{\Gamma(a+b)}\,.
\ee 
The second equality in \eqref{eq:IR4} can be shown explicitly by using the identity
\be \label{eq:GxG1-x}
	\Gamma (x) \, \Gamma (1 - x) = \frac{\pi}{\sin (\pi \, x)}\,.
\ee
Therefore,
\begin{subequations} \label{eq:CMR}
\begin{align} 
	\CI^{(4)} & = B (\alpha_{u^{}_\text{L}}, \alpha_{s^{}_\text{L}}) \, \sin ( \pi \, \alpha_{s^{}_\text{R}} ) \, B( \alpha_{s^{}_\text{R}}, \alpha_{t^{}_\text{R}}) \label{eq:CMR1} \\[2pt]
	& = B \bigl(\alpha_{s^{}_\text{L}}, \alpha_{u^{}_\text{L}} ) \, \sin ( \pi \, \alpha_{u^{}_{\text{R}}} ) \, B ( \alpha_{u^{}_{\text{R}}}, \alpha_{t^{}_\text{R}} )\,. \label{eq:CMR2}
\end{align}
\end{subequations}
The same result in \eqref{eq:CMR1} can be confirmed by evaluating $\CI^{(4)}$ in \eqref{eq:I4original} directly as in \cite{blumenhagen2012basic}.\,\footnote{Note that the integral in \eqref{eq:I4original} is defined to be over the complex plane with $d^2 z_2 = dx \, d\tilde{y}$\,, which is different from the convention used in \cite{blumenhagen2012basic}. This leads to an extra factor 1/2 in our result.} Recall that, to derive this result, we fixed $\zeta$ while performing the integral over $\xi$\,. Alternatively, if we switch the order of integration by first performing the integral over $\zeta$ with $\xi$ fixed, then 
\begin{subequations} \label{eq:CML}
\begin{align} 
	\CI^{(4)} & = B (\alpha_{u^{}_{\text{R}}}, \alpha_{s^{}_\text{R}} ) \, \sin (\pi \, \alpha_{s^{}_\text{L}} ) \, B ( \alpha_{s^{}_\text{L}}, \alpha_{t^{}_\text{L}} ) \label{eq:CML1} \\[2pt]
	& = B (\alpha_{s^{}_\text{R}}, \alpha_{u^{}_{\text{R}}} ) \, \sin ( \pi \, \alpha_{u^{}_\text{L}} ) \, B ( \alpha_{u^{}_\text{L}}, \alpha_{t^{}_\text{L}} )\,. \label{eq:CML2}
\end{align}
\end{subequations}
It is also straightforward to show that the expressions in \eqref{eq:CMR} and \eqref{eq:CML} are equivalent by repetitively using \eqref{eq:GxG1-x}. For example, applying \eqref{eq:GxG1-x} to the beta functions in \eqref{eq:CMR1} gives
\begin{align}
\begin{split}
	& \quad B (\alpha_{u^{}_\text{L}}, \alpha_{s^{}_\text{L}}) \, \sin ( \pi \, \alpha_{s^{}_\text{R}} ) \, B( \alpha_{s^{}_\text{R}}, \alpha_{t^{}_\text{R}}) \\[2pt]
	& = \frac{\sin (\pi \, \alpha_{s^{}_\text{R}}) \, \sin (\pi \, \alpha_{t^{}_\text{L}}) \, \sin (\pi \, \alpha_{u^{}_{\text{R}}})}{\sin (\pi \, \alpha_{s^{}_\text{L}}) \, \sin (\pi \, \alpha_{t^{}_\text{R}}) \, \sin (\pi \, \alpha_{u^{}_\text{L}})} \, B (\alpha_{u^{}_{\text{R}}}, \alpha_{s^{}_\text{R}} ) \, \sin ( \pi \, \alpha_{s^{}_\text{L}} ) \, B ( \alpha_{s^{}_\text{L}}, \alpha_{t^{}_\text{L}} )\,.	
\end{split}
\end{align}
From \eqref{eq:KL12-KR12}, we find 
\be \label{eq:sinrel}
	\sin \bigl( \tfrac{1}{2} \, \alpha' \pi \, K_{\text{R}i} \cdot K_{\text{R}j} \bigr) = (-1)^{n_i w_j + n_j w_i} \sin ( \tfrac{1}{2} \, \alpha' \pi \, K_{\text{L}i} \cdot K_{\text{L}j} )\,,
\ee
which implies
\begin{align}
\begin{split}
	& \quad B (\alpha_{u^{}_\text{L}}, \alpha_{s^{}_\text{L}}) \, \sin ( \pi \, \alpha_{s^{}_\text{R}} ) \, B( \alpha_{s^{}_\text{R}}, \alpha_{t^{}_\text{R}}) \\[2pt]
	& = (-1)^{n_1 \, (w_2 + w_3 + w_4) \, + \, (n_2 + n_3 + n_4) \, w_1} \, B (\alpha_{u_{\text{R}}}, \alpha_{s^{}_\text{R}} ) \, \sin ( \pi \, \alpha_{s^{}_\text{L}} ) \, B ( \alpha_{s^{}_\text{L}}, \alpha_{t^{}_\text{L}} ) \\[2pt]
	& = (-1)^{- 2 \, n_1 \, w_1} \, B (\alpha_{u_{\text{R}}}, \alpha_{s^{}_\text{R}} ) \, \sin ( \pi \, \alpha_{s^{}_\text{L}} ) \, B ( \alpha_{s^{}_\text{L}}, \alpha_{t^{}_\text{L}} ) \\[2pt]
	& = B (\alpha_{u_{\text{R}}}, \alpha_{s^{}_\text{R}} ) \, \sin ( \pi \, \alpha_{s^{}_\text{L}} ) \, B ( \alpha_{s^{}_\text{L}}, \alpha_{t^{}_\text{L}} )\,.
\end{split}
\end{align}
This shows explicitly that \eqref{eq:CMR1} and \eqref{eq:CML1} are equivalent. 

\section{Open String Amplitude and KLT Relation} \label{sec:osKLT}

The decomposition of the four closed string tachyon amplitude into the left-moving and right-moving parts in \eqref{eq:coo} is already in form the same as the standard KLT relation that relates closed string amplitudes to open string amplitudes.  The exceptions here include the cocycle factors and that $\CI_\text{L}^{(4)}$ and $\CI_\text{R}^{(4)}$ depend on $K_{\text{L}i}$ and $K_{\text{R}i}$\,, respectively. The kinematic data $K_{\text{L}i}$ and $K_{\text{R}i}$ are different from each other when the winding number $w_i$ is nonzero. As we shall see the expressions of $\CI_{\text{L}, \, \text{R}}^{(4)}$ in \eqref{eq:ILR} $\sim$ \eqref{eq:IR2} correspond to    open string amplitudes. This will be the  case by introducing appropriate D-brane configurations, which we discuss in detail below. 

\subsection{Winding Open Strings} \label{sec:wos}

Consider a stack of D-branes that extend in all the noncompactified directions and are localized in    the compactified $X$-direction. We require that the $q$-th brane is located at 
\be
    x_q = x_0 + q \, L \quad \text{mod} \quad 2\pi R\,,
        \qquad
    q \in \mathbb{Z}\,,
\ee
in the compactified circle, such that the $q$-th and $(q+1)$-th branes are separated by $L$\,. See Figure \ref{fig:dbconfig} for an illustration, where $L = 2 \pi \tilde{R}$ is fixed. 

We now take the worldsheet to be a strip, with $\tau \in \mathbb{R}$ and $\sigma \in (0, \pi)$\,. Consider open strings that satisfy the Dirichlet boundary condition in the compactified $X$-direction and the Neumann boundary condition in directions along the D-branes. We fix the ends of a given open string at two of the D-branes -- say, the $q$-th and $(q+n)$-th D-brane, with $n \in \mathbb{Z}$ -- separated by a distance $|n| L$\,.\,\footnote{Note that $|n| L$ can be larger than the circumference of the compactified circle, in which case the physical distance between the two D-branes along the circle can be smaller than $|n| L$\,.}
The compactified worldsheet field $X(\tau, \sigma)$ takes on the following mode expansion on the strip:
\be \label{eq:Xopen}
	X (\tau, \sigma) = x_q + \frac{1}{\pi} \, ( n L + 2 \pi R \, w) \, \sigma - \sqrt{2 \, \alpha'} \sum_{m \neq 0} \frac{\alpha_m}{m} \, e^{-m \tau} \sin (m \, \sigma)\,.
\ee 
Here, $x_q$ is the location of the $q$-th D-brane and $x_{q+n} = x_q + n L$ is the location of the $(q+n)$-th D-brane. 
The winding number $w$ denotes how many additional loops the open string winds around the compactified circle. Note that, when $x_{q+n} = x_q$\,, the open string ends on the same D-brane. The worldsheet field $X(\tau, \sigma)$ in \eqref{eq:Xopen} satisfies the Dirichlet boundary conditions 
\be
	X (\tau, 0) = x_q\,, 
		\qquad
	X (\tau, \pi) = x_{q+n}\,. 
\ee

Under the change of coordinates introduced in \eqref{eq:zzdef},
\be
	z = e^{\tau + i \sigma}\,,
		\qquad
	\overline{z} = e^{\tau - i \sigma}\,,
\ee 
the worldsheet is mapped to the upper half plane with
$X (z\,, \overline{z}) = X_\text{L} (z) + X_\text{R} (\overline{z})$
and
\begin{subequations}
\begin{align}
	X_\text{L} (z) & = x^{}_\text{L} - \frac{i}{2\pi} (n L + 2 \pi R \, w) \ln z - i \sqrt{\frac{\alpha'}{2}} \sum_{m \neq 0} \frac{\alpha_m}{m} \frac{1}{z^m}\,, \\[2pt]
	X_\text{R} (\overline{z}) & = x^{}_\text{R} + \frac{i}{2\pi} (n L + 2 \pi R \, w) \ln \overline{z} + i \sqrt{\frac{\alpha'}{2}} \sum_{m \neq 0} \frac{\alpha_m}{m} \frac{1}{{\overline{z}}^m}\,,
\end{align} 
\end{subequations}
with $x_q = x_\text{L} + x_\text{R}$\,.
The boundary of the wordsheet is at $z = \overline{z}$\,. 

The OPE between two $X$'s on the disk that satisfy the Dirichlet boundary conditions is
\be \label{eq:osXX}
	X (z_1) \, X (z_2) \sim - \frac{\alpha'}{2} \lr \ln |z_1 - z_2|^2 - \ln |z_1 - \overline{z}_2|^2 \rr,
\ee
It then follows that the OPEs between $X_\text{L}$'s and $X_\text{R}$'s are
\begin{subequations}
\begin{align}
	X_\text{L} (z_1) \, X_\text{L} (z_2) & \sim - \frac{\alpha'}{2} \ln z_{12}\,, 
		&
	X_\text{L} (z_1) \, X_\text{R} (\overline{z}_2) & \sim \frac{\alpha'}{2} \ln (z_1 - \overline{z}_2)\,, \\[2pt]
	X_\text{R} (\overline{z}_1) \, X_\text{R} (\overline{z}_2) & \sim - \frac{\alpha'}{2} \ln \overline{z}_{12}\,,
		&
	X_\text{R} (\overline{z}_1) \, X_\text{L} (z_2) & \sim \frac{\alpha'}{2} \ln (\overline{z}_1 - z_2)\,.
\end{align}
\end{subequations}
The OPEs on the boundary can be readily derived by setting $z_i = \overline{z}_i = y_i$\,; the OPE in \eqref{eq:osXX} vanishes on the boundary. 

The total winding number of the open string can be non-integer, given by
\be \label{eq:wnos}
	W = \frac{1}{2 \pi R} \oint_{\CC_\text{o}} \lr dz \, \p_z X_{\text{L}} + d \overline{z} \, \p_{\overline{z}} X_{\text{R}} \rr = \frac{n L}{2\pi R} + w\,, 
\ee
where the contour $\CC$ for an open string vertex operator inserted at $y$ on the real axis traverses counterclockwise along a small semi-circle centered at $y$ in the upper half plane. We refer to $n L / (2\pi R)$ in \eqref{eq:wnos} as the \emph{fractional} winding number of the open string, which means that the open string winds over a fraction of the compactified circle. From \eqref{eq:wnos} we define the winding number operator
\be
	\hat{W} = \frac{1}{2 \pi R} \oint_{\CC_\text{o}} \lr dz \, \p_z X + d \overline{z} \, \p_{\overline{z}} X \rr.
\ee
Similarly, the momentum operator in the compactified direction is
\be
    \hat{p} = \frac{1}{2\pi\alpha'} \oint_{\CC_\text{o}} \bigl( dz \, \p_z X_\text{L} - d\overline{z} \, \p_{\overline{z}} X_\text{R} \bigr)\,.
\ee
We are interested in the open string tachyon vertex operator that corresponds to the eigenstate of $\hat{W}$ with the eigenvalue $W$ in \eqref{eq:wnos}. Such a vertex operator takes the form
\be \label{eq:Co}
	\CV_\text{o} (y) = g_\text{o} : \! e^{i Q_\text{L} \cdot \mathbb{Y}_{\text{L}} (y) + i Q_\text{R} \cdot \mathbb{Y}_{\text{R}} (y)} \! :\,,
\ee
where $y \in \mathbb{R}$ and
\begin{subequations} \label{eq:osKLR}
\begin{align} 
	\mathbb{Y}^M_{\text{L,R}} \equiv \mathbb{X}^M_{\text{L,R}} \big|_{z = \overline{z}}\,,
		\qquad
	Q_{\text{L}}^M & = (k^\mu, \, W R / \alpha')\,, \label{eq:osKLRa} \\[2pt]
	Q_{\text{R}}^M & = (k^\mu, \, - W R / \alpha')\,. \label{eq:osKLRb}
\end{align}
\end{subequations}
The vertex operator $\CV_\text{o}$ corresponds to an eigenstate of $\hat{W}$ with eigenvalue $W$ and of $\hat{p}$ with eigenvalue $p = 0$\,, consistent with that open strings carry zero momentum in the direction satisfying the Dirichlet boundary condition.  
We recall that $\mathbb{X}^M = (X^\mu, X)$\,, with $X^\mu$ satisfying the Neumann boundary condition and $X$ satisfying the Dirichlet boundary condition. 
The BRST invariance of $\CV_\text{o}$ requires
\be
	- k^2 = \frac{W^2 R^2}{{\alpha'}^2} - \frac{1}{\alpha'}\,,
\ee
which gives the on-shell condition. 

\begin{figure}[t!]
\centering
\begin{minipage}{4.5cm}
\begin{tikzpicture}
    \draw [darkblue,line width = .25mm,domain=-150:-30] plot ({2.15*cos(\x)}, {2.15*sin(\x)});
    \draw[line width = .65 mm] (0,0) circle (10ex);
    \draw[line width = .55mm] (0,1.2) --(0,2.4);
    \draw[line width = .55mm] (1.2*0.866025,-1.2*.5) --(2.4*0.866025,-2.4*.5);
    \draw[line width = .55mm] (-1.2*0.866025,-1.2*.5) --(-2.4*0.866025,-2.4*.5);
    \draw[dotted] (0,0)--(0,-1.85);
    \draw[black] (-0.3,-.75) node {\scalebox{0.9}{$R$}};
    \draw[darkblue] (2.15*.866,-2.15*.5) node {$\bullet$};
    \draw[darkblue] (-2.15*.866,-2.15*.5) node {$\bullet$};
    \draw [line width = 0.65mm,domain=-29.5:89.5] plot ({10ex*cos(\x)}, {10ex*sin(\x)});
    \draw[black] (1.86,1.1) node {\scalebox{0.9}{$L$}};
    \draw [->, dotted, line width= .25mm,domain=36:87] plot ({2.1*cos(\x)}, {2.1*sin(\x)});
    \draw [->, dotted,line width= .25mm,domain=24:-26] plot ({2.1*cos(\x)}, {2.1*sin(\x)});
\end{tikzpicture}
\end{minipage}
\hspace{2cm}
\begin{minipage}{4.5cm}
\begin{tikzpicture}[thick]
    \draw [darkblue,line width= .25mm,domain=-150:-30] plot ({2*cos(\x)}, {2*sin(\x)});
    \draw [darkblue,line width= .25mm,domain=-270:-152] plot ({1.97*cos(\x)}, {2.15*sin(\x)});
    \draw [darkblue,line width= .25mm,domain=-390.5:-270] plot ({2.23*cos(\x)}, {2.15*sin(\x)});
    \draw [darkblue,line width= .25mm,domain=-510:-390] plot ({2.2*cos(\x)}, {2.27*sin(\x)});
    \draw[line width = .65 mm] (0,0) circle (10ex);
    \draw[line width = .55mm] (0,1.2) --(0,2.4);
    \draw[line width = .55mm] (1.2*0.866025,-1.2*.5) --(2.4*0.866025,-2.4*.5);
    \draw[line width = .55mm] (-1.2*0.866025,-1.2*.5) --(-2.4*0.866025,-2.4*.5);
    \draw[dotted] (0,0)--(0,-1.5-.3);
    \draw[black] (-0.3,-.75) node {\scalebox{0.9}{$R$}};
    \draw[darkblue] (2*.866,-2*.5) node {$\bullet$};
    \draw[darkblue] (-2.2*.87,-2.3*.48) node {$\bullet$};
\end{tikzpicture}
\end{minipage}
\caption{\small The circle of radius $R$ represents the compactified $X$-direction and the ticks represent a series of D-branes transverse to the compactified circle, equidistantly positioned along $X$ at $x_q = x_0 + q \, L$\,, $q = 0, \, 1, \, 2, \, \ldots$\,, as shown in Figure \ref{fig:dbconfig}. 
The diagrams above depict the special case when $2 \pi R / L = 3$\,, which is rational. As a result, maximally, there are only three transverse branes in the circle. Note that these diagrams also represent the D-brane configuration when $L$ takes other values, for example, $L = 4 \pi R / 3$\,. 
An open string, shown in blue, can stretch between two D-branes as in the left diagram, or as in the right diagram by making an integer number of wrappings around the compactified direction. Generically, when $2 \pi R / L$ is irrational, there is no maximal number of transverse branes in the circle.}
\label{fig:compactified}
\end{figure}
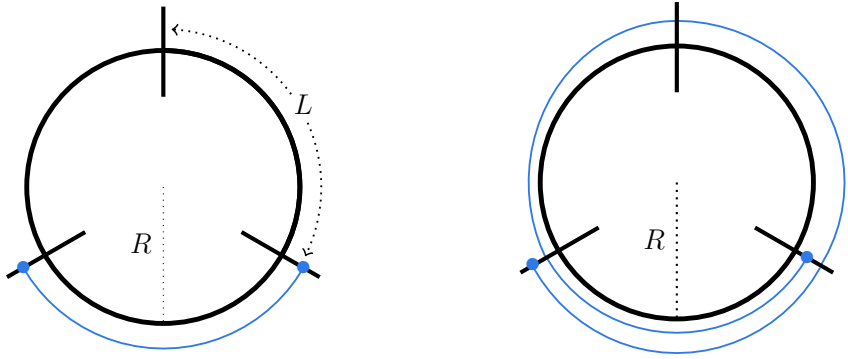

As a final remark, we note that, in the degenerate case when $2 \pi R / L$ is rational, there can be D-branes coinciding with each other in the compactified direction.  See Figure \ref{fig:compactified} for a special case with $2\pi R / L = 3$\,. When $2\pi R / L$ is irrational, none of the D-branes will coincide after compatification and the branes densely fill the circle. 

\subsection{Four Open String Tachyon Amplitudes} \label{sec:KLTfirst}

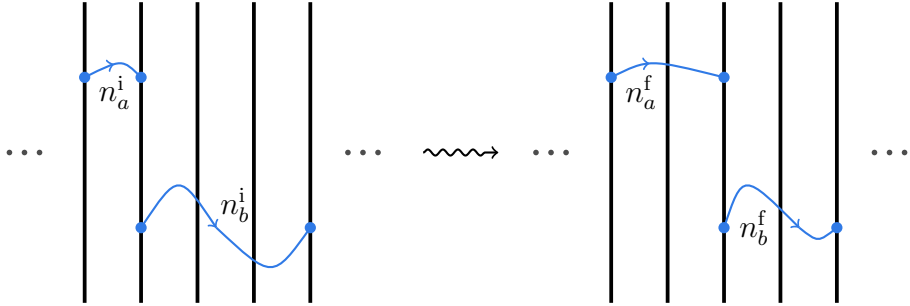
\begin{figure}[b!]
\centering
\begin{tikzpicture}[thick]
    \tikzstyle{vertex}=[circle,fill=black!70, minimum size=1pt,inner sep=.75pt]
    \node[vertex]at ( -1-3.5 , 0) {};
    \node[vertex]at ( -1-3.5+.2 , 0) {};
    \node[vertex]at ( -1-3.5+.4 , 0) {};
    \draw[black,line width= .5mm] (-.5-3,-2) -- (-.5-3,2) ;
    \draw[black,line width= .5mm] (0.25-3,-2) -- (0.25-3,2) ;
    \draw[black,line width= .5mm] (1-3,-2) -- (1-3,2) ;
    \draw[black,line width= .5mm] (1.75-3,-2) -- (1.75-3,2) ;
    \draw[black,line width= .5mm] (2.5-3,-2) -- (2.5-3,2) ;
    \tikzstyle{vertex}=[circle,fill=black!70, minimum size=1pt,inner sep=.75pt]
    \node[vertex]at ( 0 , 0) {};
    \node[vertex]at ( 0+.2 , 0) {};
    \node[vertex]at ( 0+.4 , 0) {};
    \tikzstyle{vertex}=[circle,fill=darkblue!100, minimum size=1.5pt,inner sep=1.5pt]
    \node[vertex]at ( -3.5 , 1) {};
    \node[vertex]at ( -2.75 , 1) {};
    \node[vertex]at ( -2.75 , -1) {};
    \node[vertex]at ( -.5 , -1) {};
    \draw[darkblue, decoration={markings, mark=at position 0.5 with
        {\arrow{>}}}, postaction={decorate}] (-3.5,1).. controls (-3,1.25) .. (-2.75,1);
    \draw[darkblue, decoration={markings, mark=at position 1 with
        {\arrow{>}}}, postaction={decorate}] (-2.75,-1).. controls (-2.25,-.25) .. (-1.75,-1);
    \draw[darkblue] (-1.75,-1).. controls (-1,-1.7) .. (-.5,-1);
    \draw[->,decorate,decoration={snake,amplitude=.4mm,segment length=2mm,post length=1mm}] (1,0) -- (2, 0);
    \tikzstyle{vertex}=[circle,fill=black!70, minimum size=1pt,inner sep=.75pt]
    \node[vertex]at ( -1-3.5+7 , 0) {};
    \node[vertex]at ( -1-3.5+.2+7 , 0) {};
    \node[vertex]at ( -1-3.5+.4 +7, 0) {};
    \draw[black,line width= .5mm] (-.5-3+7,-2) -- (-.5-3+7,2) ;
    \draw[black,line width= .5mm] (0.25-3+7,-2) -- (0.25-3+7,2) ;
    \draw[black,line width= .5mm] (1-3+7,-2) -- (1-3+7,2) ;
    \draw[black,line width= .5mm] (1.75-3+7,-2) -- (1.75-3+7,2) ;
    \draw[black,line width= .5mm] (2.5-3+7,-2) -- (2.5-3+7,2) ;
    \tikzstyle{vertex}=[circle,fill=black!70, minimum size=1pt,inner sep=.75pt]
    \node[vertex]at ( 0+7 , 0) {};
    \node[vertex]at ( 0+.2+7 , 0) {};
    \node[vertex]at ( 0+.4+7 , 0) {};
    \tikzstyle{vertex}=[darkblue, circle,fill=darkblue!100, minimum size=1.5pt,inner sep=1.5pt]
    \node[vertex]at ( -3.5+7 , 1) {};
    \node[vertex]at ( -2+7 , 1) {};
    \node[vertex]at ( -2+7 , -1) {};
    \node[vertex]at ( -.5+7 , -1) {};
    \draw[darkblue, decoration={markings, mark=at position 0.35 with
        {\arrow{>}}}, postaction={decorate}] (-3.5+7,1).. controls (-3+7,1.25) .. (-2+7,1);
    \draw[darkblue,decoration={markings, mark=at position 1 with
        {\arrow{>}}}, postaction={decorate}] (-2+7,-1).. controls (-1.75+7,-.25) .. (-1+7,-1);
    \draw[darkblue] (-1+7,-1).. controls (-.75+7,-1.2) .. (-.5+7,-1);
    \node at (-3.1,0.8) {$n_a^\text{i}$};
    \node at (-3.1+7,0.8) {$n_a^\text{f}$};
    \node at (-1.5,-0.7) {$n_b^\text{i}$};
    \node at (-1.5+6.9,-1) {$n_b^\text{f}$};
\end{tikzpicture}
\caption{\small The vertical lines represent D-branes separated by the same distance $L$\,, and the blue wavy lines represent fundamental strings with their ends residing on different D-branes. The picture on the left represents the initial state and the picture on the right represents the final state of the scattering process. During the scattering process, the two strings join into one at the second D-brane from the left; then, the single intermediate string splits at the third D-brane into two. This picture captures the conservation law $n_1 + n_2 + n_3 + n_4 = 0$\,, with $n_1 = 1$\,, $n_2 = 3$\,, and $n_3 = n_4 = -2$\,. The incoming strings on the left carry the initial quantum numbers $n_a^\text{i} = n_1$ and $n_b^\text{i} = n_2$\,. The outgoing strings on the right carry the quantum numbers $n_a^\text{f} = - n_3$ and $n_b^\text{f} = - n_4$\,. The amplitude for this type of scattering process is given by \eqref{eq:opsam}. The total number of D-branes involved in the scattering process is equal to $n_1 + n_2 + 1 = 5$\,.}
\label{fig:openStringDbrane}
\end{figure}
\begin{figure}[t!]
\centering
\begin{tikzpicture}[thick]
    \tikzstyle{vertex}=[circle,fill=black!70, minimum size=1pt,inner sep=.75pt]
    \node[vertex]at ( -1-3.5 , 0) {};
    \node[vertex]at ( -1-3.5+.2 , 0) {};
    \node[vertex]at ( -1-3.5+.4 , 0) {};
    \draw[black,line width= .5mm] (-.5-3,-2) -- (-.5-3,2) ;
    \draw[black,line width= .5mm] (0.25-3,-2) -- (0.25-3,2) ;
    \draw[black,line width= .5mm] (1-3,-2) -- (1-3,2) ;
    \draw[black,line width= .5mm] (1.75-3,-2) -- (1.75-3,2) ;
    \draw[black,line width= .5mm] (2.5-3,-2) -- (2.5-3,2) ;
    \tikzstyle{vertex}=[circle,fill=black!70, minimum size=1pt,inner sep=.75pt]
    \node[vertex]at ( 0 , 0) {};
    \node[vertex]at ( 0+.2 , 0) {};
    \node[vertex]at ( 0+.4 , 0) {};
    \tikzstyle{vertex}=[circle,fill=darkblue!100, minimum size=1.5pt,inner sep=1.5pt]
    \node[vertex]at ( -3.5 , 1) {};
    \node[vertex]at ( -2.75 , 1) {};
    \node[vertex]at ( -2 , -1) {};
    \node[vertex]at ( -.5 , -1) {};
    \draw[darkblue, decoration={markings, mark=at position 0.5 with
        {\arrow{>}}}, postaction={decorate}] (-3.5,1).. controls (-3,1.25) .. (-2.75,1);
    \draw[darkblue, decoration={markings, mark=at position .35 with
        {\arrow{<}}}, postaction={decorate}] (-2,-1).. controls (-1,-1.7) .. (-.5,-1);
    \draw[->,decorate,decoration={snake,amplitude=.4mm,segment length=2mm,post length=1mm}] (1,-0) -- (2, -0);
    \draw[darkblue, decoration={markings, mark=at position .27 with
        {\arrow{>}}}, postaction={decorate}] (-2.4,0) circle (1.4ex);
    \tikzstyle{vertex}=[circle,fill=black!70, minimum size=1pt,inner sep=.75pt]
    \node[vertex]at ( -1-3.5+7 , 0) {};
    \node[vertex]at ( -1-3.5+.2+7 , 0) {};
    \node[vertex]at ( -1-3.5+.4 +7, 0) {};
    \draw[black,line width= .5mm] (-.5-3+7,-2) -- (-.5-3+7,2) ;
    \draw[black,line width= .5mm] (0.25-3+7,-2) -- (0.25-3+7,2) ;
    \draw[black,line width= .5mm] (1-3+7,-2) -- (1-3+7,2) ;
    \draw[black,line width= .5mm] (1.75-3+7,-2) -- (1.75-3+7,2) ;
    \draw[black,line width= .5mm] (2.5-3+7,-2) -- (2.5-3+7,2) ;
    \tikzstyle{vertex}=[circle,fill=black!70, minimum size=1pt,inner sep=.75pt]
     	 \node[vertex]at ( 0+7 , 0) {};
    \node[vertex]at ( 0+.2+7 , 0) {};
    \node[vertex]at ( 0+.4+7 , 0) {};
    \tikzstyle{vertex}=[circle,fill=darkblue!100, minimum size=1.5pt,inner sep=1.5pt]
    \node[vertex]at ( -3.5+7 , 1) {};
    \node[vertex]at ( -2+7 , 1) {};
    \node[vertex]at ( -2.75+7 , -1) {};
    \node[vertex]at ( -.5+7 , -1) {};
    \draw[darkblue, decoration={markings, mark=at position 0.35 with 
        {\arrow{>}}}, postaction={decorate}] (-3.5+7,1).. controls (-2.6+7,1.25) .. (-2+7,1);
    \draw[darkblue, decoration={markings, mark=at position 1 with
        {\arrow{<}}}, postaction={decorate}] (-2.75+7,-1).. controls (-2+7,-.25) .. (-1.5+7,-1);
    \draw[darkblue] (-1.5+7,-1).. controls (-1+7,-1.7) .. (-.5+7,-1);
    \node at (-3.1,0.75) {$\tilde{n}_a^\text{i}$};
    \node at (-1.6,-0.9) {$\tilde{n}_b^\text{i}$};
    \node at (-1.5+7,-0.45) {$\tilde{n}_b^\text{f}$};
    \node at (-3.1+7,0.7) {$\tilde{n}_a^\text{f}$};
\end{tikzpicture}
\caption{\small Scattering  between two open strings that do not end on the same D-brane necessarily involve asymptotic closed string states. The figure with closed strings realizes the same conservation law for $n_i$ in Figure \ref{fig:openStringDbrane} as follows. The incoming strings on the left carry the initial quantum numbers $\tilde{n}_a^\text{i} = n_1$ and $\tilde{n}_b^\text{i} = n_3$\,. The outgoing strings on the right carry the final quantum numbers $\tilde{n}_a^\text{f} = -n_4$ and $\tilde{n}_b^\text{f} = -n_2$\,. Nevertheless, the scattering process considered in this figure requires inserting an external closed string vertex operator, and is thus \emph{not} computed by \eqref{eq:opsam}.}
\label{fig:closedstringintermediate}
\end{figure}
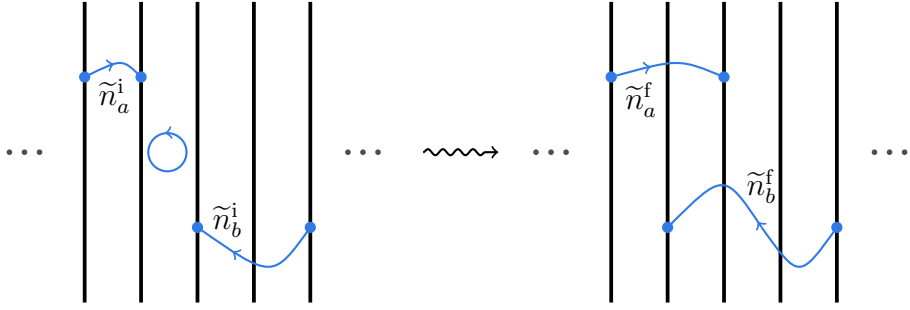

We are ready to compute open string scattering amplitudes with the open strings ending on the D-brane configuration introduced in \S\ref{sec:wos}. We are interested in the following open string amplitude of four open string tachyons inserted along the real axis in the complex plane, with the ends of each open string attached to the D-branes:
\begin{align} \label{eq:opsam}
	\CA_\text{o} (1, 2, 3, 4) & = e^{- \chi \Phi_0} \left \langle \, \prod_{i=1}^4 \! : \! c(\mathbb{Y}_i) \, \CV_\text{o} (\mathbb{Y}_i) \! : \right \rangle_{\!\!\text{D}_2}
		\qquad%
\begin{minipage}{2.8cm}
\begin{tikzpicture}
	\draw (0,0) circle (0.7cm); 
	\filldraw (-0.707*0.7,0.707*0.7) circle (0.5mm);
	\filldraw (0.707*0.7,0.707*0.7) circle (0.5mm);
	\filldraw (-0.707*0.7,-0.707*0.7) circle (0.5mm);
	\filldraw (0.707*0.7,-0.707*0.7) circle (0.5mm);
	\node at (-0.907*0.7,0.807*0.7) {$1\,$};
	\node at (0.907*0.7,0.807*0.7) {$\,2$};
	\node at (0.907*0.7,-0.807*0.7) {$\,3$};
	\node at (-0.907*0.7,-0.807*0.7) {$4\,$};
\end{tikzpicture}
\end{minipage}
\end{align}
The Euler characteristic $\chi = 1$ for the disk and we have specified the permutation. We apply the PSL(2, $\mathbb{R}$) symmetry to fix $y_1 = 0$\,, $y_3 = 1$\,, and $y_4 = \infty$\,. Then,
\begin{align} \label{eq:Npointosa}
\begin{split}
	\CA_\text{o} (1, 2, 3, 4) 
	\propto \frac{i}{\alpha'} \, (2\pi)^{25} \delta^{(25)} (k_1 + \cdots + k_4) \, \delta_{W_1 + \cdots + W_4, \, 0} \, \CM_\text{o} (1, 2, 3 , 4)\,.
\end{split}
\end{align}
Here,
\be \label{eq:CMo}
	\CM_\text{o} (1, 2, 3, 4) = \int_{0}^1 dy_2 \, |y_2|^{2 \, \alpha' K_1 \cdot K_2} \, |1 - y_2|^{2 \, \alpha' K_2 \cdot K_3}\,,
\ee
with 
\be \label{eq:osK}
    K_i^M = (k^\mu_i\,, W_i \, R / \alpha')\,.
\ee
We have dropped a proportionality constant that depends on the coupling constant $g_\text{o}$\,. In general, there can also be color factors from Chan-Paton charges, which we drop when the KLT relation is concerned.  The total winding number $W_i$ defined in \eqref{eq:wnos} is
\be \label{eq:tw}
	W_i = \frac{n_i L}{2 \pi R} + w_i\,,
		\qquad
	n_i\,, w_i \in \mathbb{Z}\,.
\ee
The conservation law of $W_i$ is imposed by the Kronecker symbol in \eqref{eq:Npointosa}, which gives
\be \label{eq:conlawW}
    \frac{L}{2\pi R} \sum_{i} n_i + \sum_{i} w_i = 0\,.
\ee
In the case when $L / (2\pi R)$ is irrational, \eqref{eq:conlawW} further implies
\be \label{eq:splitconlawW}
    \sum_i n_i = \sum_i w_i = 0\,.
\ee
In the degenerate case when $L / (2 \pi R)$ is rational, it is still possible to choose the definitions of $n_i$ and $w_i$ such that the split of the conservation law \eqref{eq:conlawW} in \eqref{eq:splitconlawW} is satisfied. We therefore require that \eqref{eq:splitconlawW} hold in the rest of the paper.

Note that the amplitude  in \eqref{eq:Npointosa} involves only open strings, which can only join or split on a D-brane. See Figure \ref{fig:openStringDbrane} for an illustration. The total number of D-branes that are involved in the scattering process is given by $n_+ + 1$\,, where $n_+$ is the sum of $n_i > 0$ associated with the incoming closed strings. See Figure \ref{fig:css} and \ref{fig:dbconfig}.
In contrast, scatterings between open strings that do not occur on the same D-brane necessarily involve closed strings. One such example is given in Figure \ref{fig:closedstringintermediate}. These scattering processes with asymptotic closed string states are not captured by the open string amplitudes that we compute here and are irrelevant in the KLT relation that we consider in this paper.

Intriguingly, if we take $L = 2 \, \pi \tilde{R}$\,, with $\tilde{R} = \alpha' / R$ the radius of the T-dual circle over which the $X$-direction is compactified, we find that $K_i^M$ in \eqref{eq:osK} becomes
\be \label{eq:PiM}
	K_{i}^M = \lr k^\mu_i\,, \, \frac{n_i L}{2\pi\alpha'} + \frac{w_i R}{\alpha'} \rr = \lr k^\mu_i\,, \, \frac{n_i}{R} + \frac{w_i R}{\alpha'} \rr,
\ee
which is identical to $K_{\text{L}i}$ defined for closed strings in \eqref{eq:KLMdef}, and $n_i$ can be interpreted as the KK number for the closed string tachyon.\footnote{If $w_i$ in $K_{i}^M$ is identified with the negative of the closed string winding number, $K_{i}$ in \eqref{eq:PiM} is mapped to $K_{\text{R}i}$ defined in \eqref{eq:KRMdef}.}
For the open string amplitude in \eqref{eq:CMo} to match the expression in \eqref{eq:ILR1}, we have to make the following rescalings of $\alpha'$\,, $R$\,, and $L$\,: 
\be \label{eq:rescalingaR}
	\alpha' \rightarrow \frac{\alpha'}{4}\,,
		\qquad
	R \rightarrow \frac{R}{4}\,,
	    \qquad
	L \rightarrow \frac{L}{4} = \frac{\pi \alpha'}{2 R}\,.
\ee
Here, the rescaling of $R$ and $L$ essentially uniformly rescales the geometry of the system. Under the rescalings in \eqref{eq:rescalingaR}, the open string amplitude \eqref{eq:CMo} becomes
\be \label{eq:ILo}
	\CM_\text{L} (1, 2, 3, 4) \equiv \CM_\text{o} (1, 2, 3, 4)_{\alpha' \rightarrow \frac{\alpha'}{4}, \, R \rightarrow \frac{R}{4}, \, L \rightarrow \frac{L}{4}} = \CI_\text{L}^{(4)}\,, 
\ee
where $\CI_\text{L}^{(4)}$ is given in \eqref{eq:ILR1}. 

Similarly, for different permutations of vertex operators and appropriate choices of which D-branes the open strings end at, we find the following open string amplitude interpretation for the quantities in \eqref{eq:IR1} and \eqref{eq:IR2},
\begin{subequations} 
\begin{align}
	\CI_\text{R}^{(4)} & = 2 \, i \sin \bigl( \tfrac{1}{2} \, \pi \alpha' K_{\text{R}1} \cdot K_{\text{R}2} \bigr) \, \CM_\text{R} (1, 3, 4, 2)
		\qquad\quad\!%
		\label{eq:IRo}
\begin{minipage}{2.8cm}
\begin{tikzpicture}
	\draw (0,0) circle (0.7cm); 
	\filldraw (-0.707*0.7,0.707*0.7) circle (0.5mm);
	\filldraw (0.707*0.7,0.707*0.7) circle (0.5mm);
	\filldraw (-0.707*0.7,-0.707*0.7) circle (0.5mm);
	\filldraw (0.707*0.7,-0.707*0.7) circle (0.5mm);
	\node at (-0.907*0.7,0.807*0.7) {$1\,$};
	\node at (0.907*0.7,0.807*0.7) {$\,3$};
	\node at (0.907*0.7,-0.807*0.7) {$\,4$};
	\node at (-0.907*0.7,-0.807*0.7) {$2\,$};
\end{tikzpicture}
\end{minipage} \\[2pt]
	& = 2 \, i \sin \bigl( \tfrac{1}{2} \, \pi \alpha' K_{\text{R}2} \cdot K_{\text{R}3} \bigr) \, \CM_{\text{R}} (1, 3, 2, 4)
		\qquad\quad\!%
		\label{eq:IRo2}
\begin{minipage}{2.8cm}
\begin{tikzpicture}
	\draw (0,0) circle (0.7cm); 
	\filldraw (-0.707*0.7,0.707*0.7) circle (0.5mm);
	\filldraw (0.707*0.7,0.707*0.7) circle (0.5mm);
	\filldraw (-0.707*0.7,-0.707*0.7) circle (0.5mm);
	\filldraw (0.707*0.7,-0.707*0.7) circle (0.5mm);
	\node at (-0.907*0.7,0.807*0.7) {$1\,$};
	\node at (0.907*0.7,0.807*0.7) {$\,3$};
	\node at (0.907*0.7,-0.807*0.7) {$\,2$};
	\node at (-0.907*0.7,-0.807*0.7) {$4\,$};
\end{tikzpicture}
\end{minipage}
\end{align}
\end{subequations}
We have defined
\be \label{eq:MRdef}
	\CM_\text{R} (i_1, i_2, i_3, i_4) \equiv \CM_\text{o} (i_1, i_2, i_3, i_4)_{\alpha' \rightarrow \frac{\alpha'}{4}, \, R \rightarrow \frac{R}{4}, \, L \rightarrow \frac{L}{4}, \, w_i \rightarrow - w_i}\,. 
\ee
The second identity in \eqref{eq:IRo2} comes from the open string relation
\be \label{eq:BCJ}
    \sin \bigl( 2 \pi \alpha' P_{1} \cdot P_{2} \bigr) \, \CM_{\text{o}} (1, 3, 4, 2) = \sin \bigl( 2 \pi \alpha' P_{2} \cdot P_{3} \bigr) \, \CM_{\text{o}} (1, 3, 2, 4)\,,
\ee
which is the stringy generalization \cite{BjerrumBohr:2009rd, Stieberger:2009hq} of the Bern-Carraso-Johansson (BCJ) relation in field theories \cite{Bern:2008qj}. However, in the case of nonzero winding, the field theory limit $\alpha' \rightarrow 0$ does not converge, as winding states do not have a conventional QFT interpretation. 

\subsection{Four-Point KLT Relation} \label{sec:fpKLT}

We are finally ready to write down the first KLT relation for winding strings. Plugging \eqref{eq:ILo} and \eqref{eq:IRo} into \eqref{eq:Mc4I4} and \eqref{eq:coo}, we find the following KLT relation between closed string amplitude $\CM_\text{c}$ and open string amplitudes $\CM_\text{L}$ and $\CM_\text{R}$\,:
\begin{subequations} \label{eq:Mc1234}
\begin{align}
	& \quad \CM_\text{c} (1, 2, 3, 4) =\notag \\[2pt]
	& = - \frac{1}{\alpha'} \, C (1, 2, 3, 4) \, \sin \bigl( \tfrac{1}{2} \, \pi \alpha' K_{\text{R}1} \cdot K_{\text{R}2} \bigr) \, \CM_{\text{L}} (1, 2, 3, 4) \, \CM_{\text{R}} (2, 1, 3, 4)  \label{eq:KLT1} \\[2pt]
	& = - \frac{1}{\alpha'} \, C (1, 2, 3, 4) \, \sin \bigl( \tfrac{1}{2} \, \pi \alpha' K_{\text{R}2} \cdot K_{\text{R}3} \bigr) \, \CM_{\text{L}} (1, 2, 3, 4) \, \CM_{\text{R}} (1, 3, 2, 4)\,, \label{eq:KLT2}
\end{align}
\end{subequations}
where $C$ denotes contribution  from the cocycle factors, with
\be
    C (i_1\,, i_2\,, i_3\,, i_4) \equiv \exp \Bigl( i \pi \, {\sum}^4_{{\substack{p, \, q = 1 \\ p < q}}} \, n_{i_p} \, w_{i_q} \Bigr)\,.
\ee
Here, \eqref{eq:KLT1} matches \eqref{eq:CMR1} and \eqref{eq:KLT2} matches \eqref{eq:CMR2}. We now show explicitly that \eqref{eq:Mc1234} is invariant under switching ``L" and ``R." First, note that the closed string amplitude is unchanged upon permuting the vertex operators. Starting with \eqref{eq:KLT1} and performing the permutation $(1, 2, 3, 4) \rightarrow (3, 2, 4, 1)$\,, we have
\begin{align}
\begin{split}
	& \quad \CM_\text{c} (1, 2, 3, 4) = \CM_\text{c} (3, 2, 4, 1) \\[2pt]
	& = - \frac{1}{\alpha'} \, C(3, 2, 4, 1) \, \sin \bigl( \tfrac{1}{2} \, \pi \alpha' K_{\text{R}3} \cdot K_{\text{R}2} \bigr) \, \CM_{\text{L}} (3, 2, 4, 1) \, \CM_{\text{R}} (2, 3, 4, 1)\,.
\end{split}
\end{align}
Using
\begin{align}
    	C(3, 2, 4, 1) 
    	& = (-1)^{n_3 w_2 - n_2 w_3} \, C(1, 2, 3, 4)\,,
\end{align}
and applying \eqref{eq:sinrel}, 
we find
\begin{align}
\begin{split}
	& \quad \CM_\text{c} (1, 2, 3, 4) \\[2pt]
	& = - \frac{1}{\alpha'} \, C(1, 2, 3, 4) \, \sin \bigl( \tfrac{1}{2} \, \pi \alpha' K_{\text{L}2} \cdot K_{\text{L}3} \bigr) \, \CM_{\text{R}} (1, 2, 3, 4) \, \CM_{\text{L}} (1, 3, 2, 4)\,,
\end{split}
\end{align}
which is \eqref{eq:KLT2} but with ``L" and ``R" switched, and it matches \eqref{eq:CML2}. Similarly, we can show that \eqref{eq:KLT2} is equal to
\be
	- \frac{1}{\alpha'} \, C(1, 2, 3, 4) \, \sin \bigl( \tfrac{1}{2} \, \pi \alpha' K_{\text{L}1} \cdot K_{\text{L}2} \bigr) \, \CM_{\text{R}} (1, 2, 3, 4) \, \CM_{\text{L}} (2, 1, 3, 4)\,,
\ee
which is \eqref{eq:KLT1} but with ``L" and ``R" switched, and it matches \eqref{eq:CML1}.

From the above four tachyon amplitudes, we make the following general observations of the KLT relation for winding strings, which are also valid for amplitudes involving massless and massive   vertex operators. The KLT relation in \eqref{eq:Mc1234} is for string amplitudes in spacetime with a spatial direction compactified over a circle of radius $R$\,. The KLT relation maps the closed string amplitude to a sum of the products of a pair of open string amplitudes with the open strings ending on an array of D-branes transverse to the compactified direction. The D-branes are positioned equidistantly along the spatial circle, with the consecutive ones separated by $L = 2 \pi \tilde{R}$\,, where $\tilde{R} = \alpha' / R$ is the T-dual radius. For a closed string that carries winding number $w_i$ and KK number $n_i$\,,\footnote{For tachyons we have $n_i w_i = 0$\,, which implies that at least one of $n_i$ and $w_i$ is zero. However, when  other vertex operators are considered, the quantity $n_i w_i$ is not necessarily zero.} the corresponding open string wraps $w_i$ times around the compactified circle. In addition, excluding the $w_i$ times that it wraps around the circle, the open string is stretched between the $q$-th and $(q+n_i)$-th branes.\,\footnote{Since the series of D-branes is wrapped around the compactified circle, the shortest separation along the circle between the ends of the open string might span over fewer than $n_i-1$ D-branes. For example, in the diagram on the right in Figure \ref{fig:compactified}, if $w_i = 0$\,, the open string starts at the $q$-th brane and ends at the $(q+4)$-th brane in the series, which implies that $n_i = 4$\,. However, the shortest separation between the ends of the open string along the compactified circle is just between two D-branes.} Using \eqref{eq:wnos}, we find that the closed string momentum $n_i / R$ along the compactified direction is mapped to the fractional winding number (defined in \eqref{eq:wnos})
\be
	\frac{L}{2\pi R} \, n_i = \frac{\tilde{R}}{R} \, n_i = \frac{\alpha'}{R^2} \, n_i
\ee
of the open strings. The number of D-branes that are involved in the scattering between open strings is $n_+ + 1$\,, with $n_+$ the sum of $n_i > 0$ associated with the incoming closed strings. See Figure \ref{fig:css} $\sim$ \ref{fig:osamp} for illustrations.

There is an equivalent interpretation for the open string amplitudes in the T-dual frame, where the open strings satisfy   Neumann boundary conditions in the compactified direction. In this case, the open strings can carry nonzero momenta but only zero winding. The KK number of a closed string is mapped to the KK number of the associated open string, while the winding number of a closed string is mapped to a constant gauge field, which gives rise to a shift in the open string momentum: it gives rise to fractional momentum. The gauge field $A$ is associated with a $U(n_+ + 1)$ Chan-Paton charge, where $n_+ + 1$ is   the number of D-branes. Since we take $A$ to be constant, one can diagonalize it to be
\be
    A = \frac{1}{2\pi \tilde{R}} \, \text{diag} (\theta_1\,, \ldots, \theta_{n_+ + 1})\,,
        \qquad
    \theta_q = \frac{q L}{2 \pi R}\,,
\ee
breaking the gauge group to $U(1)^{n_+ + 1}$. Note that $\theta_q$ is T-dual to the location of the $q$-th D-brane. The field $A$ is pure gauge, but a Wilson loop that winds around the compactified circle of dual radius $\tilde{R}$ gains a phase factor of the form 
\be
    \exp \lr i \oint dX \, A_q \rr = e^{i \theta_q},
\ee 
leading to a shift in the momentum $\tilde{n}_i / \tilde{R}$ along the compactified $X$-direction. For a string in the Chan-Paton state $| q+\tilde{w}_i\,,\, q \rangle$\,, its canonical momentum in $X$ is
\be \label{eq:Pidual}
    \tilde{K}^M_{i} = \lr k^\mu_i\,, \, \frac{\tilde{n}_i + \theta_{q+\tilde{w}_i} - \theta_{q}}{\tilde{R}} \rr = \lr k^\mu_i\,, \, \frac{\tilde{n}_i}{\tilde{R}} + \frac{\tilde{w}_i \tilde{R}}{\alpha'}\rr.
\ee
Here, $\tilde{n}_i = w_i$ and $\tilde{w}_i = n_i$\,, with $w_i$ ($n_i$) the winding (momentum) number of the corresponding closed string. 

The above observations hold for any higher point amplitudes among general vertex operators discussed in \S\ref{sec:hovo}. We prove the general KLT relation for winding strings in the next section. We also note that all such constructions straightforwardly generalize to the Narain compactification, in which more than one spatial direction  is compactified. In this more   case, we need to consider D$p$-branes that are transverse to multiple compactified directions.

\section{KLT Relation for Higher-Point Amplitudes} \label{sec:hpa}

In this section, we derive the KLT relation for higher-point amplitudes, involving winding closed string vertex operators that take the general form as discussed in \S\ref{sec:hovo}. 

\subsection{Higher-Point Closed String Amplitude}

We start with an $\CN$-point tree-level closed string amplitude, which, similar to the four-point amplitude in \eqref{eq:A4}, takes the following general form: 
\be
	\CA_\text{c}^{(\CN)} \propto (2\pi)^{25} \, \delta^{(25)} \Bigl( {\sum}_{i=1}^\CN k_i \Bigr) \, \delta_{{\sum}_{i=1}^\CN n_i\,, \, 0} \, \delta_{{\sum}_{i=1}^\CN w_i\,, \, 0} \, \bigl( i \CM^{(\CN)}_\text{c} \bigr)\,,
\ee
where
\begin{align} 
\begin{split}
	& \CM_\text{c} (1\,, \ldots, \, \CN) = \frac{1}{{\alpha'}^{\CN-3}} \,  |z_1 - z_{\CN-1}|^2 \, |z_1 - z_\CN|^2 \, |z_{\CN-1} - z_\CN|^2 \\[2pt]
	& \times \! \lr \prod_{i=2}^{\CN-2} \int_{\mathbb{C}} d^2 z_i \rr F(z_1\,, \ldots, z_\CN\,; \, \overline{z}_1\,, \ldots, \overline{z}_\CN) \prod_{\substack{j, k = 1 \\ j < k}}^{\CN} e^{i \pi w_j n_k}  \, z^{\frac{1}{2} \alpha' K_{\text{L}j} \cdot K_{\text{L}k}}_{jk} \, \overline{z}^{\frac{1}{2} \alpha' K_{\text{R}j} \cdot K_{\text{R}k}}_{jk}.
\end{split}
\end{align}
We defined $z_{ij} \equiv z_i - z_j$\,. Here, the factor $F$ comes from contracting the derivative terms in the vertex operators that for example take the form of \eqref{eq:gvo1} and \eqref{eq:gvo2}. The factor $F$ is single valued. Fixing $z_1 = 0$\,, $z_{\CN-1} = 1$\,, and $z_\CN = \infty$\,, we find
\begin{align} \label{eq:CMcCN}
\begin{split}
	& \quad \CM_\text{c} (1\,, \ldots, \, \CN) \\[2pt]
	& = \frac{1}{{\alpha'}^{\CN-3}} \, C(1\,, \ldots, \, \CN) 
	\lr \prod_{\substack{i = 2}}^{\CN-2} \int_{\mathbb{C}} d^2 z_i \rr F(z_2\,, \ldots, z_{\CN-2}\,; \, \overline{z}_2\,, \ldots, \overline{z}_{\CN-2}) \\[2pt]	
	& \hspace{4cm} \times \! \prod_{\substack{i, \, j, \, k = 2 \\ j < k}}^{\CN-2} z_i^{\frac{1}{2} \alpha' K_{\text{L}1} \cdot K_{\text{L}i}} \, (1-z_i)_{\phantom{i}}^{\frac{1}{2} \alpha' K_{\text{L}(\CN -1)} \cdot K_{\text{L}i}} \, z^{\frac{1}{2} \alpha' K_{\text{L}j} \cdot K_{\text{L}k}}_{kj} \times \\[-6pt]
	& \hspace{5.55cm} \overline{z}_i^{\frac{1}{2} \alpha' K_{\text{R}1} \cdot K_{\text{R}i}} \, (1 - \overline{z}_i)^{\frac{1}{2} \alpha' K_{\text{R}(\CN-1)} \cdot K_{\text{R}i}} \, \overline{z}^{\frac{1}{2} \alpha' K_{\text{R}j} \cdot K_{\text{R}k}}_{kj},
\end{split}
\end{align}
where the cocycle factor $C(1\,, \ldots, \, \CN)$ is defined by
\be
    C (i_{1}\,, \ldots, i_\CN) \equiv \exp \Bigl( i \pi \, {\sum}_{\substack{p, \, q = 1 \\ p < q}}^\CN \, n_{i_p} \, w_{i_q} \Bigr)\,.
\ee
We have abbreviated $F(z_1\,, \ldots, z_\CN\,; \, \overline{z}_1\,, \ldots, \overline{z}_\CN) \big|_{z_1 = 0, \, z_{\CN-1} = 1, \, z_\CN = \infty}$ by dropping the arguments $z_1$\,, $z_{\CN-1}$\,, and $z_\CN$\,, as well as their complex conjugates.

We write schematically the $i$-th closed string vertex operators as
\begin{align} \label{eq:gvo}
\begin{split}
	& \sum_{\substack{k_1 + \cdots k_p = N \\ \ell_1 + \cdots \ell_q = \tilde{N}}} \!\!\! \epsilon^{(i)}_{I_1 \cdots I_{p} \, J_1 \cdots J_q} \, \Bigl[ e^{\tfrac{1}{2} \, i \pi R \, w_i \, (\hat{p}^{}_{\text{L}} + \hat{p}^{}_{\text{R}})} \Bigr]_{\CC_i} :\! \p^{\,k_1} \mathbb{X}_{\text{L}i}^{I_1} \cdots \p^{\,k_p} \mathbb{X}_{\text{L}i}^{I_p} \,\, {\overline{\p}}{}^{\,\ell_1} \mathbb{X}_{\text{R}i}^{J_{1}} \cdots {\overline{\p}}{}^{\,\ell_q} \mathbb{X}_{\text{R}i}^{J_{q}} \\[-14pt]
	& \hspace{9cm} \times e^{i K_{\text{L}i} \cdot \mathbb{X}_{\text{L}i} (z) + i K_{\text{R}i} \cdot \mathbb{X}_{\text{R}i} (\overline{z})} \!:.
\end{split}
\end{align}
With appropriate choices of the polarization tensors $\epsilon^{(i)}$\,, this general operator reduces to the vertex operators considered in \eqref{eq:gvo1} and \eqref{eq:gvo2}. In general, the tensors $\epsilon^{(i)}$ satisfy various conditions by imposing BRST invariance on the vertex operator. The dispersion relation and the level matching condition for the state corresponding to the vertex operator \eqref{eq:gvo} are in form the same as the expressions in \eqref{eq:gdrlm}.
In the following calculation, we employ the standard trick in \cite{Kawai:1985xq} and rewrite the polarization tensor as
\be \label{eq:replace1}
	\epsilon^{(i)}_{I_1 \cdots I_{p} \, J_1 \cdots J_q} \rightarrow \epsilon^{\text{L}i}_{I_1 \cdots I_p} \, \epsilon^{\text{R}i}_{J_1 \cdots J_q}\,.
\ee
Under this replacement, the $F$-factor in \eqref{eq:CMcCN} decomposes as
\be \label{eq:replaceF}
	F(z_2\,, \ldots, z_{\CN-2}\,; \, \overline{z}_2\,, \ldots, \overline{z}_{\CN-2}) \rightarrow F_\text{L} (z_2\,, \ldots, z_{\CN-2}) \, F_\text{R} (\overline{z}_2\,, \ldots, \overline{z}_{\CN-2})\,,
\ee
where $F_{\text{L}}$ ($F_{\text{R}}$) contains polarization factor $\epsilon^{\text{L}i}$ ($\epsilon^{\text{R}i}$). The new factors $F_{\text{L}, \text{R}}$ will be interpreted as factors in open string amplitudes coming from contracting derivative terms in open string vertex operators. Applying the replacement \eqref{eq:replaceF} to \eqref{eq:CMcCN}, we find
\begin{align} \label{eq:Mcrep}
\begin{split}
	& \quad \CM_\text{c} (1\,, \ldots, \, \CN) \\[2pt]
	& \rightarrow \frac{1}{{\alpha'}^{\CN-3}} \,  C(1\,, \ldots, \, \CN) \lr \prod_{\substack{i = 2}}^{\CN-2} \int_{\mathbb{C}} d^2 z_i \rr F_\text{L}(z_2\,, \ldots, z_{\CN-2}) \, F_{\text{R}} (\overline{z}_2\,, \ldots, \overline{z}_{\CN-2}) \\[2pt]
	& \hspace{4cm} \, \times \! \prod_{\substack{i, \, j, \, k = 2 \\ j < k}}^{\CN-2} z_i^{\frac{1}{2} \alpha' K_{\text{L}1} \cdot K_{\text{L}i}} \, (1-z_i)_{\phantom{i}}^{\frac{1}{2} \alpha' K_{\text{L}(\CN-1)} \cdot K_{\text{L}i}} \, z^{\frac{1}{2} \alpha' K_{\text{L}j} \cdot K_{\text{L}k}}_{kj} \times \\[-6pt]
	& \hspace{5.5cm} \,\, \overline{z}_i^{\frac{1}{2} \alpha' K_{\text{R}1} \cdot K_{\text{R}i}} \, (1 - \overline{z}_i)^{\frac{1}{2} \alpha' K_{\text{R}(\CN-1)} \cdot K_{\text{R}i}} \, \overline{z}^{\frac{1}{2} \alpha' K_{\text{R}j} \cdot K_{\text{R}k}}_{kj}.
\end{split}
\end{align}
Note that $F_\text{L,\,R}$ are single-valued functions that do not contribute any branch points. Collecting terms multilinear in $\epsilon^{\text{L}i}$ and $\epsilon^{\text{R}i}$ and replacing their product with $\epsilon^{(i)}$, we recover the original amplitude \eqref{eq:CMcCN}. In the following, we will take $\CM_\text{c}$ to denote the expression in \eqref{eq:Mcrep}, with the splitting of the polarization factor taken into account implicitly. 

\subsection{Splitting the Worldsheet Integrals}

It is a standard exercise to perform the integrals over the complex plane such that it splits into two sets of contour integrals for the left- and right-movers, respectively \cite{Kawai:1985xq}, bearing caveats concerning the cocycle factor. For completeness, we will review the complex integral and articulate the differences. 

First, make the change of variables with $z_ i = x_i + i \, \tilde{y}_i$ for real variables $x_i$ and $\tilde{y}_i$\,. Then, we promote the integrals over $\tilde{y}_i$ to be in the complex plane, with the branch points at $i \, x_i$\,, $- i (1 - x_i)$\,, and $\tilde{y}_j + i (x_i - x_j)$\,.  Deform the contour for the $\tilde{y}_i$\,-integral that runs along the real axis by rotating it counterclockwise by $\frac{\pi}{2} - \epsilon$\,, $\epsilon \rightarrow 0^+$\,. This is done by a second change of variables,
\be
	\tilde{y}_i = \exp \bigl[ i (\tfrac{\pi}{2} - \epsilon) \bigr] \, y_i \approx i \, y_i + \epsilon \, y_i\,.
\ee
We require that such deformations of the contours are done simultaneously for all the $\tilde{y}_i$\,-integrals, in order to ensure that none of the branch points are crossed. Define
\be \label{eq:defquan}
	\xi_i \equiv x_i + y_i\,,
		\qquad
	\zeta_i \equiv x_i - y_i\,,
		\qquad
	\delta_i \equiv \epsilon \, y_i = \tfrac{1}{2} \, \epsilon \, (\xi_i - \zeta_i)\,,
\ee
in terms of which we rewrite \eqref{eq:Mcrep} as
\be \label{eq:tMczx}
	\CM_\text{c} (1\,, \ldots, \, \CN) = \frac{1}{{\alpha'}^{\CN-3}} \, C (1\,, \ldots, \, \CN) \lr \prod_{\substack{i = 2}}^{\CN-2}  \int_{\mathbb{R}} d\zeta_i \int_{\mathbb{R}} d\xi_i \rr O_\text{L} \, O_\text{R}\,,
\ee
where
\begin{subequations}
\begin{align}
	O_\text{L} = F_\text{L} & \prod_{\substack{j = 2}}^{\CN-2} \, (\zeta_j + i \, \delta_j)^{\frac{1}{2} \alpha' K_{\text{L}1} \cdot K_{\text{L}j}} \, (1 - \zeta_j - i \, \delta_j)_{\phantom{i}}^{\frac{1}{2} \alpha' K_{\text{L}(\CN-1)} \cdot K_{\text{L}j}} \notag \\[2pt]
	\times & \! \prod_{\substack{k, \, \ell = 2 \\ k < \ell}}^{\CN-2}  \bigl[ \zeta_\ell - \zeta_k + i \, (\delta_\ell - \delta_k) \bigr]^{\frac{1}{2} \alpha' K_{\text{L}k} \cdot K_{\text{L}\ell}} \,, \label{eq:OL1} \\[6pt]
	O_\text{R} = F_\text{R} & \prod_{\substack{j = 2}}^{\CN-2} (\xi_j - i \, \delta_j)^{\frac{1}{2} \alpha' K_{\text{R}1} \cdot K_{\text{R}j}} \, (1 - \xi_j + i \, \delta_j)_{\phantom{i}}^{\frac{1}{2} \alpha' K_{\text{R}(\CN-1)} \cdot K_{\text{R}j}} \notag \\[2pt]
		\times & \! \prod_{\substack{k, \, \ell = 2 \\ k < \ell}}^{\CN-2}  \bigl[ \xi_\ell - \xi_k - i \, (\delta_\ell - \delta_k) \bigr]^{\frac{1}{2} \alpha' K_{\text{R}k} \cdot K_{\text{R}\ell}} \,.
\end{align}
\end{subequations}
We now perform the integrals over $\xi_i$ for fixed $\zeta_i$\,, $i = 2, \ldots, \, \CN-2$\,. 
First, we define the permutation $\sigma (2, \ldots, \, \CN-2)$ of the particle numbers such that the fixed values of $\zeta_i$'s are ordered as 
\be \label{eq:orderingzeta}
	\zeta_{\sigma (2)} < \zeta_{\sigma (3)} < \cdots < \zeta_{\sigma (\CN-2)}\,. 
\ee
In the case when $\zeta_{\sigma(2)} \in (- \infty\,, 0)$\,, the branch points for the $\xi_{\sigma(2)}$\,-integral are at 
\begin{subequations}
\begin{align}
	i \, \delta_{\sigma(2)} \big|_{\xi_{\sigma(2)} = 0} & = \tfrac{i}{2} \, \epsilon \, (- \zeta_{\sigma(2)})\,, \\[2pt]
	1 + i \, \delta_{\sigma_{(2)}} \big|_{\xi_{\sigma(2)} = 1} & = 1 + \tfrac{i}{2} \, \epsilon \, \bigl( 1 - \zeta_{\sigma(2)} \bigr)\,, \\[4pt]
	\xi_{\sigma(i)} + i \, (\delta_{\sigma(2)} - \delta_{\sigma(i)}) \big|_{\xi_{\sigma(2)} = \xi_{\sigma(i)}} & = \xi_{\sigma(i)} + \tfrac{i}{2} \, \epsilon \, (\zeta_{\sigma(i)} - \zeta_{\sigma(2)})\,, \quad i > 2\,.
\end{align}
\end{subequations}
All these branch points reside in the upper half plane. Completing the contour in the lower half plane, we find that the contribution from $\zeta_{\sigma(2)} \in (-\infty\,, 0)$ to the amplitude $\CM_\text{c}$ is zero. Similarly, when $\zeta_{\sigma(\CN-2)} \in (1\,, \infty)$\,, all the branch points for the $\xi_{\sigma(\CN-2)}$\,-integral are in the lower half plane. Therefore, the contribution from $\zeta_{\sigma(\CN-2)} \in (1, \infty)$ to $\CM_\text{c}$ is also zero. Therefore, the integrals in \eqref{eq:tMczx} are only nonzero when
\be
	0 < \zeta_{\sigma(2)} < \cdots < \zeta_{\sigma(\CN-2)} < 1\,,
\ee
which partitions up the associated integration region from 0 to 1.
As a result, \eqref{eq:tMczx} becomes
\begin{align} \label{eq:Mcsigma}
\begin{split}
	& \quad \CM_\text{c} (1\,, \ldots, \, \CN) = \CM_\text{c} \bigl(1\,, \sigma(2\,, \ldots, \, \CN-2)\,, \, \CN-1\,, \, \CN \bigr) \\[2pt]
	& = \lr \frac{i}{2 \alpha'} \rr^{\CN-3} \!\! C \bigl(1\,, \sigma(2\,, \ldots, \, \CN-2)\,, \, \CN-1\,, \, \CN \bigr) \int_{\Delta} \lr \prod_{\substack{i = 2}}^{\CN-2} d\zeta_i \int_{\mathbb{R}} d\xi_i \rr O_\text{L} \, O_\text{R}\,,
\end{split}
\end{align}
with the integration domain 
\be \label{eq:Deltadomain}
    \Delta = \{ \zeta_2\,, \ldots, \zeta_{\CN-2} \big| 0 < \zeta_{\sigma(2)} < \cdots < \zeta_{\sigma(\CN-2)} < 1 \}\,,
\ee
and
\begin{subequations} \label{eq:Mcnew}
\begin{align}
	O_\text{L} = F_\text{L} & \prod_{
	j = 2}^{\CN-2} \zeta_j^{\frac{1}{2} \alpha' K_{\text{L}1} \cdot K_{\text{L}j}} \, (1 - \zeta_j)_{\phantom{i}}^{\frac{1}{2} \alpha' K_{\text{L}(\CN-1)} \cdot K_{\text{L}j}} \notag \\[2pt]
	\times & \! \prod_{\substack{k, \, \ell = 2 \\ k < \ell}}^{\CN-2} (\zeta_{\sigma(\ell)} - \zeta_{\sigma(k)})^{\frac{1}{2} \alpha' K_{\text{L}\sigma(k)} \cdot K_{\text{L}\sigma(\ell)}} \,, \label{eq:OL2} \\[6pt]
	O_\text{R} = F_\text{R} & \prod_{\substack{j = 2}}^{\CN-2} (\xi_j - i \, \delta_j)^{\frac{1}{2} \alpha' K_{\text{R}1} \cdot K_{\text{R}j}} \, (1 - \xi_j + i \, \delta_j)_{\phantom{i}}^{\frac{1}{2} \alpha' K_{\text{R}(\CN-1)} \cdot K_{\text{R}j}} \notag \\[2pt]
	\times & \! \prod_{\substack{k, \, \ell = 2 \\ k < \ell}}^{\CN-2}  \bigl[ \xi_{\sigma(\ell)} - \xi_{\sigma(k)} - i \, (\delta_{\sigma(\ell)} - \delta_{\sigma(k)}) \bigr]^{\frac{1}{2} \alpha' K_{\text{R}\sigma(k)} \cdot K_{\text{R}\sigma(\ell)}} \,. \label{eq:McnewOR}
\end{align}
\end{subequations}
When $k < \ell$\,, 
for a fixed $\xi_{\sigma(\ell)}$\,, the $\xi_{\sigma(k)}$\,-integral has a branch point at 
\be
    \xi_{\sigma(\ell)} + \tfrac{i}{2} \, \epsilon \bigl( \zeta_{\sigma(\ell)} - \zeta_{\sigma(k)} \bigr)\,,
\ee
which resides in the upper half plane if the $\xi_{\sigma(k)}$\,-contour traverses the real axis. This implies that, in order to avoid branch points, we need to shift the contour associated with the $\xi_{\sigma(\ell)}$\,-integral to be slightly above the contour associated with the $\xi_{\sigma(k)}$\,-integral in the complex plane \cite{BjerrumBohr:2010hn}. 

We now elaborate on the factors
\be \label{eq:zetaik}
	(\zeta_{\sigma(\ell)} - \zeta_{\sigma(k)})^{\frac{1}{2} \alpha' K_{\text{L}\sigma(k)} \cdot K_{\text{L}\sigma(\ell)}}\,, 
		\qquad
	k < \ell\,,
\ee 
that appear in \eqref{eq:OL2}. When $\sigma (\ell) < \sigma (k)$\,, this factor ultimately comes from 
\be \label{eq:zetakl}
	(z_{\sigma(k)} - z_{\sigma(\ell)})^{\frac{1}{2} \alpha' K_{\text{L}\sigma(k)} \cdot K_{\text{L}\sigma(\ell)}} 
\ee 
in \eqref{eq:Mcrep}. To pass from \eqref{eq:zetakl} to \eqref{eq:zetaik}, an extra phase factor that is a function of $K_{\text{L}i}$ appears. The same reordering for $\overline{z}_{i}$'s in \eqref{eq:Mcrep} also introduces a similar phase factor, which is a function of $K_{\text{R}i}$\,. In the case when there is zero winding, we have $K_{\text{L}i} = K_{\text{R}i}$ and there is no extra phase factor introduced after simultaneously reordering $z_i$'s and $\overline{z}_i$'s. This is, however, not the case when winding modes are present: The extra phase factors from reordering $z_i$'s and $\overline{z}_i$'s do \emph{not} exactly cancel when $K_{\text{L}i} \neq K_{\text{R}i}$\,. The remainder phase factor can be absorbed into the cocycle factor $C (1\,, \ldots, \, \CN)$ in \eqref{eq:tMczx}, which turns the cocycle factor into $C \bigl( 1\,, \, \sigma(2\,, \ldots, \, \CN-2)\,, \, \CN-1\,, \, \CN \bigr)$ in \eqref{eq:Mcsigma}, where the arguments are permuted. The above observation is summarized by the following identity:
\begin{align}
\begin{split}
	& \quad C (1, \ldots, \, \CN) \prod_{\substack{k, \, \ell = 2 \\ k < \ell}}^{\CN-2} (z_\ell - z_k)^{\frac{1}{2} \alpha' K_{\text{L}k} \cdot K_{\text{L}\ell}} \, (\overline{z}_\ell - \overline{z}_k)^{\frac{1}{2} \alpha' K_{\text{R}k} \cdot K_{\text{R}\ell}} \\[2pt]
	& = C \bigl(1\,, \sigma(2\,, \ldots, \, \CN-2)\,, \, \CN-1\,, \, \CN \bigr) \\[2pt]
	& \hspace{2.35cm} \times \prod_{\substack{k, \, \ell = 2 \\ k < \ell}}^{\CN-2}  (z_{\sigma(\ell)} - z_{\sigma(k)})^{\frac{1}{2} \alpha' K_{\text{L}\sigma(k)} \cdot K_{\text{L}\sigma(\ell)}}
	\, (\overline{z}_{\sigma(\ell)} - \overline{z}_{\sigma(k)})^{\frac{1}{2} \alpha' K_{\text{R}\sigma(k)} \cdot K_{\text{R}\sigma(\ell)}}.
\end{split}
\end{align}  

\subsection{Higher-Point KLT Relation} \label{sec:hpKLT}

Now, the integrals over $\zeta_i$ and $\xi_i$ in \eqref{eq:Mcsigma} can be performed in the standard way as in \cite{Kawai:1985xq, Stieberger:2009hq, BjerrumBohr:2010hn}, generalizing the same procedure for four-point amplitudes in \S\ref{sec:4ptcs} to $\CN$-point amplitudes. We find the following $\CN$-point KLT relation:
\begin{align} \label{eq:Mcresult}
\begin{split}
	\CM_{\text{c}} (1, & \ldots, \, \CN) = \lr -1 \rr^{\CN-3} \, \sum_{\sigma, \, \beta, \, \gamma}  C \bigl(1\,, \sigma(2\,, \ldots, \, \CN-2)\,, \, \CN-1\,, \, \CN \bigr) \\[2pt]
	& \hspace{0.5cm} \times \mathcal{S}_\text{R} \bigl[ \gamma \circ \sigma(2\,, \ldots,\, \ell-1) \, \big | \, \sigma (2\,, \ldots,\, \ell-1)\bigr]_{K_{\text{R}1}} \\[4pt]
    & \hspace{0.5cm} \times \mathcal{S}_\text{R} \bigl[ \sigma(\ell\,, \ldots, \, \CN-2) \, \big| \, \beta \circ \sigma(\ell\,, \ldots, \, \CN-2) \bigr]_{K_{\text{R}(\CN-1)}} \\[2pt]
    & \hspace{0.5cm} \times \CM_{\text{L}} \bigl( 1\,, \, \sigma(2\,, \ldots, \, \CN-2)\,, \, \CN-1\,, \, \CN \bigr) \\[4pt]
    & \hspace{0.5cm} \times \CM_{\text{R}} \bigl( \gamma \circ \sigma(2\,, \ldots, \, \ell-1)\,, \, 1\,, \, \CN-1, \, \beta \circ \sigma (\ell\,, \ldots, \, \CN-2)\,, \, \CN \bigr)\,.
\end{split}
\end{align}
Note that $\beta$ permutes $\CN-\ell-1$ indices and $\gamma$ permutes $\ell-2$ indices. Here, we defined the open string amplitude $\CM_\text{L}$ after the rescaling \eqref{eq:rescalingaR} as
\begin{align} \label{eq:gosa}
    \CM_{\text{L}} \bigl(1, \sigma(2, \ldots, \, \CN-2), \, \CN-1, \, \CN \bigr)
    & = \! \int_{\Delta} F_\text{L} \prod_{i=2}^{\CN-2} d\zeta_i \, |\zeta_i|^{\frac{1}{2} \alpha' K_{\text{L}1} \cdot K_{\text{L}i}} |1 - \zeta_i|_{\phantom{i}}^{\frac{1}{2} \alpha' K_{\text{L}(\CN-1)} \cdot K_{\text{L}i}} \notag \\[2pt]
    & \hspace{1.1cm} \times \! \prod_{\substack{j,\, k = 2 \\ j < k}}^{\CN-2} \! |\zeta_{\sigma(k)} - \zeta_{\sigma(j)}|^{\frac{1}{2} \alpha' K_{\text{L}\sigma(j)} \cdot K_{\text{L}\sigma(k)}},
\end{align}
where the domain $\Delta$ is defined in \eqref{eq:Deltadomain}. The $i$-th open string involved in the scattering amplitude is labeled by the quantum numbers $(n_i\,, w_i)$\,. See \eqref{eq:ILo} for an example. Open string amplitudes with a different cyclic ordering are defined accordingly by permuting the integration domain and the indices in \eqref{eq:gosa}. The open string amplitude $\CM_\text{R}$ is defined analogously by replacing the subscript ``L" with ``R," with the $i$-th open string involved in the scattering amplitude labeled by the quantum numbers $(n_i\,, - w_i)$\,. See \eqref{eq:MRdef} for an example.
The momentum kernel $\mathcal{S}_\text{R}$ is defined as \cite{BjerrumBohr:2010ta}
\begin{align} 
\begin{split}
	& \quad \mathcal{S}_{\text{R}} \bigl[ i_1\,, \ldots, i_k \, | \, j_1\,, \ldots, j_k \bigr]_{P} \\[2pt]
	& = \lr \alpha' \rr^{-k} \, \prod_{t=1}^{k} \, \sin \Bigl[ \tfrac{1}{2} \, \pi \alpha' \bigl( P \cdot K_{\text{R}i_t} + \sum_{q = t+1}^k \theta(i_t\,, i_q) \, K_{\text{R}i_t} \cdot K_{\text{R}i_q} \bigr) \Bigr]\,,
\end{split}
\end{align}
which takes into account the phase factors associated to crossing a branch cut. Here, $i_1\,, \ldots, i_k$ and $j_1\,, \ldots, j_k$ are elements in $\{ 1\,, \ldots, k \}$\,. Moreover, if the ordering of $i_t$ and $i_q$ is the opposite in the ordered sets $( i_1\,, \ldots\,, i_k )$ and $( j_1\,, \ldots\,, j_k )$\,, we set $\theta(i_t\,, i_q) = 1$\,; if the ordering is the same, we set $\theta(i_t\,, i_q) = 0$\,. The KLT relation in \eqref{eq:Mcresult} shows how the $\CN$-point amplitude for closed strings with windings can be factorized as a product of open string amplitudes. The open strings satisfy the Dirichlet boundary condition in the compactified direction, with their ends positioned on an array of D-branes transverse to the compactified circle, as we have detailed in \S\ref{sec:osKLT}.

Note that \eqref{eq:Mcresult} is independent of $\ell \in \{ 3\,, \ldots, \, \CN-2 \}$\,. This is reminiscent of the four-point amplitude that we discussed earlier in \S\ref{sec:4ptcs}, which has two equivalent expressions in \eqref{eq:IR1} and \eqref{eq:IR2}, depending on whether the integration contour for the $\xi$\,-integral is deformed to the left or right. To derive the $\CN$-point amplitude \eqref{eq:Mcresult}, we have deformed $\ell - 2$ contours associated with the $\xi_{\sigma(k)}$\,-integrals, $k = 2, \, \ldots, \ell-1$\,, and the remaining $\CN - \ell - 1$ contours have been deformed to the right, without crossing any branch cuts. From the open string amplitude perspective, the existence of different ways of writing the KLT relation is due to the open string generalization of the Kleiss-Kuijf and BCJ relations \cite{BjerrumBohr:2009rd, Stieberger:2009hq, BjerrumBohr:2010hn}.\,\footnote{Also see \cite{Casali:2019ihm} for generalizations to higher genus Riemann surfaces using twisted homology.} 
The KLT relation can also be cast in equivalent forms other than \eqref{eq:Mcresult}. When $\ell = 2$\,, all the contours for the $\xi_{i}$\,-integral are deformed to the right, yielding
\begin{align} \label{eq:Mcresult1N-2}
\begin{split}
	\CM_{\text{c}} (1\,, \ldots, \, \CN) & = \lr -1 \rr^{\CN-3} \, \sum_{\rho, \, \sigma} C \bigl(1\,, \sigma(2\,, \ldots, \, \CN-2)\,, \, \CN-1\,, \, \CN \, \bigr) \\
	& \hspace{2cm} \times \mathcal{S}_\text{R} \bigl[ \sigma(2\,, \ldots, \, \CN-2) \, \big| \, \rho(2\,, \ldots, \, \CN-2) \bigr]_{K_{\text{R}(\CN-1)}} \\[2pt]
    & \hspace{2cm} \times \CM_{\text{L}} \bigl( 1\,, \, \sigma(2\,, \ldots, \, \CN-2)\,, \, \CN-1\,, \, \CN \, \bigr) \\[6pt]
    & \hspace{2cm} \times \CM_{\text{R}} \bigl( 1\,, \, \CN-1\,, \, \rho (2\,, \ldots, \, \CN-2)\,, \, \CN \, \bigr)\,.
\end{split}
\end{align}
Here, $\rho$ is the same as $\sigma$ that permutes the ordered set $(2, \ldots, \CN-2)$\,. When $\ell = \CN - 1$\,, all the contours for the $\xi_i$\,-integrals are deformed to the left, where
\begin{align} \label{eq:Mcresult1N-22}
\begin{split}
	\CM_{\text{c}} (1\,, \ldots, \, \CN) & = \lr -1 \rr^{\CN-3} \, \sum_{\rho, \, \sigma} C \bigl(1\,, \sigma(2\,, \ldots, \, \CN-2)\,, \, \CN-1\,, \, \CN \, \bigr) \\
	& \hspace{2cm} \times \mathcal{S}_\text{R} \bigl[ \rho(2\,, \ldots, \, \CN-2) \, \big| \, \sigma(2\,, \ldots, \, \CN-2) \bigr]_{K_{\text{R}1}} \\[2pt]
    	& \hspace{2cm} \times \CM_{\text{L}} \bigl( 1\,, \, \sigma(2\,, \ldots, \, \CN-2)\,, \, \CN-1\,, \, \CN \, \bigr) \\[6pt]
    	& \hspace{2cm} \times \CM_{\text{R}} \bigl( \rho (2\,, \ldots, \, \CN-2)\,, 1\,, \, \CN-1\,, \, \CN \, \bigr)\,.
\end{split}
\end{align}
The expressions in \eqref{eq:Mcresult1N-2} and \eqref{eq:Mcresult1N-22} can be thought as two limiting cases of \eqref{eq:Mcresult}.\,\footnote{There is also a different representation of KLT relations as noted in \cite{Kawai:1985xq}, which can be obtained by evaluating the integrals in \eqref{eq:tMczx} directly without performing any further deformation of the integration contours. This representation has a simple combinatorial interpretation \cite{Britto:2021prf}. In the winding case, this representation has the benefit of making the resulting KLT relation manifestly symmetric with respect to the subscripts ``L" and ``R." However, this KLT relation contains $\bigl[ (\CN-1)!/2 \bigr]^2$ terms. In contrast, the KLT relation \eqref{eq:Mcresult} is more useful in practice since it only contains $(\CN-3)! \, (\ell-2)! \, (\CN-\ell-1)!$ terms, with $2 \leq \ell \leq \CN-1$\,, which are far fewer than $\bigl[ (\CN-1)!/2 \bigr]^2$ terms when $\CN > 3$ \cite{BjerrumBohr:2010hn, Britto:2021prf}.} 

To show explicitly that these expressions are invariant under swapping the subscripts ``L" and ``R,"  we start with the expression in \eqref{eq:Mcresult1N-2} and permute the closed string vertex operators as
\begin{align} \label{eq:McRL}
	\CM_\text{c} (1\,, \ldots, \, \CN) & = \CM_\text{c} (\CN, 2\,, \, \ldots, \, \CN-2\,, 1\,, \, \CN-1) \notag \\[2pt]
	& = \lr -1 \rr^{\CN-3} \, \sum_{\rho, \, \sigma} C \bigl(\CN, \rho(2\,, \ldots, \, \CN-2)\,, 1\,, \, \CN-1 \bigr) \notag \\
	& \hspace{2.4cm} \times  \mathcal{S}_\text{R} \bigl[ \rho(2\,, \ldots, \, \CN-2) \, \big| \, \sigma(2\,, \ldots, \, \CN-2) \bigr]_{{K_{\text{R}1}}} \notag \\[2pt]
    	& \hspace{2.4cm} \times \CM_{\text{L}} \bigl( \CN, \, \rho(2\,, \ldots, \, \CN-2)\,, \, 1\,, \, \CN-1 \bigr) \notag \\[6pt]
    	& \hspace{2.4cm} \times \CM_{\text{R}} \bigl( \CN, \, 1\,, \, \sigma(2\,, \ldots, \, \CN-2)\,, \, \CN-1 \bigr) \notag \\[10pt]
	& = \lr -1 \rr^{\CN-3} \, \sum_{\rho, \, \sigma} C \bigl(\CN, \rho(2\,, \ldots, \, \CN-2)\,, 1\,, \, \CN-1 \bigr) \notag \\
	& \hspace{2.4cm} \times  \mathcal{S}_\text{R} \bigl[ \rho(2\,, \ldots, \, \CN-2) \, \big| \, \sigma(2\,, \ldots, \, \CN-2) \bigr]_{{K_{\text{R}1}}} \notag \\[2pt]
    	& \hspace{2.4cm}  \times \CM_{\text{R}} \bigl( 1\,, \, \sigma(2\,, \ldots, \, \CN-2)\,, \, \CN-1\,, \, \CN \bigr) \notag \\[6pt]
    	& \hspace{2.4cm} \times \CM_{\text{L}} \bigl( \rho(2\,, \ldots, \, \CN-2)\,, \, 1\,, \, \CN-1\,, \, \CN \bigr)\,.
\end{align}
We have swapped the dummy permutation symbol $\rho$ and $\sigma$ to get the second equality above. Furthermore, note that
\begin{align} \label{eq:Cid}
\begin{split}
	& C \bigl(\CN, \rho(2, \ldots, \, \CN-2), 1, \, \CN-1 \bigr) \,\, \mathcal{S}_\text{R} \bigl[ \rho(2, \ldots, \, \CN-2) \big| \sigma(2, \ldots, \, \CN-2) \bigr]_{{K_{\text{R}1}}} \\[2pt]
	=  \, & C \bigl(1, \sigma(2, \ldots, \, \CN-2), \, \CN-1, \, \CN \bigr) \, \mathcal{S}_\text{L} \bigl[ \rho(2, \ldots, \, \CN-2) \big| \sigma(2, \ldots, \, \CN-2) \bigr]_{{K_{\text{L}1}}},
\end{split}
\end{align}
where
\begin{align}
\begin{split}
	& \quad \mathcal{S}_{\text{L}} \bigl[ i_1\,, \cdots, i_k \, | \, j_1\,, \cdots, j_k \bigr]_{P} \\[2pt]
	& \equiv \lr \alpha' \rr^{-k} \, \prod_{t=1}^{k} \, \sin \Bigl[ \tfrac{1}{2} \, \pi \alpha' \bigl( P \cdot K_{\text{L}i_t} + \sum_{q>t}^k \theta(i_t\,, i_q) \, K_{\text{L}i_t} \cdot K_{\text{L}i_q} \bigr) \Bigr]\,.
\end{split}
\end{align}
Plugging \eqref{eq:Cid} back into \eqref{eq:McRL}, we find
\begin{align} \label{eq:McRL2}
\begin{split}
	\CM_\text{c} (1\,, \cdots, \, \CN)
	& = \lr -1 \rr^{\CN-3} \, \sum_{\rho, \, \sigma} C \bigl(1, \sigma(2, \ldots, \, \CN-2), \, \CN-1, \, \CN \bigr) \\
	& \hspace{2.4cm} \times  \mathcal{S}_\text{L} \bigl[ \rho(2\,, \cdots, \, \CN-2) \, \big| \, \sigma(2\,, \cdots, \, \CN-2) \bigr]_{{K_{\text{L}1}}} \\[2pt]
    & \hspace{2.4cm}  \times \CM_{\text{R}} \bigl( 1\,, \, \sigma(2\,, \cdots, \, \CN-2)\,, \, \CN-1\,, \CN \bigr) \\[6pt]
    & \hspace{2.4cm} \times \CM_{\text{L}} \bigl( \rho(2\,, \cdots, \, \CN-2)\,, \, 1\,, \, \CN-1\,, \CN \bigr)\,,
\end{split}
\end{align}
which is exactly the expression in \eqref{eq:Mcresult1N-22} but with the subscripts ``L" and ``R" swapped. Another way to understand this invariance under switching ``L" and ``R" is by observing that \eqref{eq:Mcresult1N-22} comes from integrating $\{\xi_i\}$ before $\{\zeta_i\}$ in \eqref{eq:tMczx}, and that \eqref{eq:McRL2} comes from integrating $\{\zeta_i\}$ before $\{\xi_i\}$ in \eqref{eq:tMczx}. Therefore, the invariance under switching ``L" and ``R" is the consequence of the invariance under commuting the orders of performing different integrals. Similarly, starting with the expression from switching ``L" and ``R" in \eqref{eq:Mcresult1N-22} and then going through the analogous procedure in \eqref{eq:McRL} $\sim $ \eqref{eq:McRL2}, we arrive at \eqref{eq:Mcresult1N-2} but with the subscripts ``L" and ``R" swapped.
It also follows immediately that the more general KLT relation in \eqref{eq:Mcresult} has to hold after switching ``L" and ``R." 

\section{Conclusions and Outlooks} \label{sec:conclusions}

In this paper we extended the KLT relation to winding  strings in  a toroidal compactification of string theory. We showed that the string amplitudes of winding closed strings factorize into quadratic products of amplitudes for open strings ending on an array of D-branes transverse to the compact directions. The winding number $w$ of a closed string is mapped to an integer winding number of an  open string wrapping around the   compactified circle; while  the momentum of a closed string is mapped to the  fractional winding number that encodes how many D-branes the open string traverses, excluding the $w$ times that the open string winds around the full circle. The general form of the KLT relation is given in \eqref{eq:Mcresult}. In Appendix \ref{app:Rpr}, we use intersection theory to write the KLT relations in a succinct form in \eqref{eq:itKLT}. 

In the case of nonzero winding, there is no field theory limit by simply sending the Regge slope $\alpha'$ to zero. However, there still is a nonsingular $\alpha' \rightarrow 0$ limit for winding  strings in the presence of a near critical Kalb-Ramond field. This type of limit  was originally applied to open strings on spacetime filling D-branes, which led to noncommutative open string (NCOS) theory \cite{Seiberg:2000ms, Gopakumar:2000na}. This NCOS limit was then realized for winding closed strings \cite{Klebanov:2000pp, Gomis:2000bd, Danielsson:2000gi} as well as winding open strings with Dirichlet boundary conditions \cite{Danielsson:2000mu, Gomis:2020fui}, leading to self-contained nonrelativistic closed and open string theory, where the string spectra enjoy a Galilean-invariant dispersion relation. It has also been shown that NCOS is T-dual to nonrelativistic open string theory \cite{Gomis:2020izd}. Recently, there has been growing interest  in nonrelativistic string theory, leading to a plethora of applications to nonrelativistic gravity and field theory (see, e.g., \cite{Andringa:2012uz, Bergshoeff:2019pij, Bergshoeff:2018yvt, Gomis:2019zyu, Gomis:2020fui, Gallegos:2019icg, Harmark:2018cdl, Harmark:2019upf, Gallegos:2020egk, Bergshoeff:2021bmc}). Amplitudes for nonrelativistic closed strings have been studied in \cite{Gomis:2000bd}, while amplitudes for nonrelativistic open strings and the associated KLT relation are still in progress \cite{nosa}. This endeavor may also enrich the current S-matrix program to include nonrelativistic field theories. 

The KLT relation for winding string amplitudes explored in this paper opens up new possibilities of generalizing other modern techniques developed for field-theoretical amplitudes to the study of winding string states. The program of mapping out structures of winding string amplitudes may bring us new insights into string theory and quantum gravity in general.

\acknowledgments

It is a pleasure to thank Freddy Cachazo and Sebastian Mizera for reading a draft and their comments. This research is supported in part by Perimeter Institute for Theoretical Physics. Research at Perimeter Institute is supported in part by the Government of Canada through the Department of Innovation, Science and Economic Development Canada and by the Province of Ontario through the Ministry of Colleges and Universities.

\newpage

\appendix

\section{KLT and Twisted Period Relation} \label{app:Rpr}

In this appendix, we review the modern interpretation of the KLT relation using intersection theory. We will see that the KLT relation for winding strings given in \S\ref{sec:hpKLT} is indeed underlay by the twisted Riemann's period relation \cite{cho1995intersection} (which generalizes Riemann's bilinear relation), and the inverse KLT kernel receives the geometrical interpretation as an intersection number between cycles in the twisted homology, defined on the moduli space.  This is a direct generalization of the results in \cite{Gaiotto:2013rk, Mizera:2017cqs} for the standard KLT relation, which we follow closely in the following discussion. For the mathematical part of the following discussion, we mostly follow \cite{aomoto:2011theory, mimachi2003intersection}.

\subsection{Twisted Cocycles and Closed String Amplitudes}

We start with the closed string side of the KLT relation. Recall the integrals for closed string amplitudes in \eqref{eq:Mcrep}, and its split form in \eqref{eq:Mcsigma}, which we choose to separate into the multi-valued function given by the Koba-Nielsen factor,
\be \label{eq:KN}
    U (z_2, \cdots, z_{\CN-2}) = \prod_{\substack{i, \, j = 1 \\ i < j}}^{\CN-1} z^{\frac{1}{2} \alpha' K_{\text{R}i} \cdot K_{\text{R}j}}_{ji},
\ee
and the following factors that are single valued:
\begin{subequations}
\begin{align}
    f_\text{L} (z_2, \cdots, z_{\CN-2}) & = F_\text{L} (z_2, \cdots, z_{\CN-2}) \prod_{\substack{i, \, j = 1 \\ i < j}}^{\CN-1} z^{\frac{1}{2} \alpha' K_{\text{L}i} \cdot K_{\text{L}j} - \frac{1}{2} \alpha' K_{\text{R}i} \cdot K_{\text{R}j}}_{ji} \notag \\[2pt]
    & = F_\text{L} (z_2, \cdots, z_{\CN-2}) \prod_{\substack{i, \, j = 1 \\ i < j}}^{\CN-1} z^{n_i w_j + n_j w_i}_{ji}\,,
        \quad
    n_i\,, w_i \in \mathbb{Z}\,;  \label{eq:fL} \\[2pt]
    f_\text{R} (\overline{z}_2, \cdots, \overline{z}_{\CN-2}) & = F_\text{R} (\overline{z}_2, \cdots, \overline{z}_{\CN-2})\,.
\end{align}
\end{subequations}
Note that $z_1 = 0$ and $z_{\CN-1} = 1$ are fixed. Using the above notation, the closed string amplitude \eqref{eq:Mcrep} can be written as
\begin{align} \label{eq:Mcffu}
    \CM_\text{c} (1, \cdots, \CN) = C (1, \cdots, \CN) & \lr \prod_{\substack{i = 2}}^{\CN-2} \int_{\mathbb{C}} d^2 z_i \rr f_\text{L} (z_2, \cdots, z_{\CN-2}) \, f_\text{R} (\overline{z}_2, \cdots, \overline{z}_{\CN-2}) \notag \\[-2pt]
    & \hspace{2.5cm} \times |U (z_2, \cdots, z_{\CN-2})|^2\,.
\end{align}
This closed string amplitude is reminiscent of the intersection number between cocycles in the twisted cohomology, which we brief in the following.

We start with the definition of the twisted cohomology \cite{matsumoto1998intersection} on the moduli space.
The singular loci of the Koba-Nielsen factor \eqref{eq:KN} are
\be \label{eq:hp}
    z_i = 0\,,
        \qquad
    z_i - 1 = 0\,,
        \qquad
    z_j - z_i = 0\,,
        \,\,
    2 \leq i < j \leq \CN-2\,.  
\ee
We define a divisor $D$ as the union of the hyperplanes in \eqref{eq:hp}. Then, the closed string moduli space is $X = \mathbb{C}^{\CN-3} \setminus D$\,. Consider the twisted logarithmic one-form of the multi-valued Koba-Nielsen factor \eqref{eq:KN} \cite{Mizera:2017cqs},
\be
    \omega \equiv d \log U = \frac{\alpha'}{2} \sum_{i = 2}^{\CN-2} E_i \, dz_i\,,
        \qquad
    E_i \equiv \sum_{\substack{j = 1 \\ j \neq i}}^{\CN-1} \frac{K_{\text{R}i} \cdot K_{\text{R}j}}{z_{ij}}\,,
\ee
and the covariant derivative $\nabla_{\!\omega} = d + \omega \wedge$ on $X$\,. We require that $\tfrac{\alpha'}{2} K_{\text{R}i} \cdot K_{\text{R}j} \in \mathbb{R} \setminus \mathbb{Z}$\,. Here, $E_i$ denotes the scattering equation \cite{Fairlie:1972zz, Cachazo:2013gna}. With respect to $\nabla_{\!\omega}$\,, we define the $(\CN-3)$-th twisted de Rham cohomology 
\be
    H^{\CN-3} (X, \nabla_{\!\omega}) = \frac{\text{Ker} \bigl(\nabla_{\!\omega}\!: \Omega^{\CN-3} (X) \rightarrow 0\bigr)}{\text{Im} \bigl(\nabla_{\!\omega}\!: \Omega^{\CN-4} (X) \rightarrow \Omega^{\CN-3} (X) \bigr)}\,,
\ee
where $\Omega^k (X)$ is the space of smooth $k$-forms on $X$\,.\footnote{A common basis for the twisted cohomology takes a logarithmic form, which is related to the Parke-Taylor factor \cite{Mizera:2017cqs, Mafra:2011nw}.} Consider the single-valued $(\CN-3)$-forms on $X$\,, 
\begin{subequations}
\begin{align}
    \varphi^{}_\text{L} (z_2, \cdots, z_{\CN-2}) & = f_\text{L} (z_2, \cdots, z_{\CN-2}) \, dz_2 \wedge \cdots \wedge dz_{\CN-2}\,, \label{eq:phiL} \\[2pt]
    \varphi^{}_\text{R} (z_2, \cdots, z_{\CN-2}) & = f_\text{R} (z_2, \cdots, z_{\CN-2}) \, dz_2 \wedge \cdots \wedge dz_{\CN-2}\,,
\end{align}
\end{subequations}
which are cocycles in $H^{\CN-3} (X, \! \nabla_{\!\omega})$\,. We now consider a pairing between $H^{\CN-3} (X, \! \nabla_{\!\omega})$ and its dual $H^{\CN-3}_\text{c} (X, \! \nabla_{\!\omega^\vee})$\,. It is important that the dual cohomology is with compact support, and hence the subscript ``c" \cite{matsumoto1998intersection}.  We keep this implicit in the following expressions. These two cohomologies are dual to each other under the pairing
\be
    \bigl\langle \varphi^{}_\text{L}\,, \varphi^{\vee}_\text{L} \bigr\rangle = \int_X \lr U \, U^\vee \rr \varphi^{}_\text{L} \wedge \varphi^{\vee}_\text{L}\,.
\ee
Defining $\omega^\vee \equiv \overline{\omega}$\,, $U^\vee \equiv \overline{U}$\,, and $\varphi_\text{L}^\vee \equiv \overline{\varphi}^{}_\text{R}$\,, where the barred quantities are complex conjugations, we find
\be \label{eq:phiLR}
    \bigl\langle \varphi^{}_\text{L}\,, \overline{\varphi}^{}_\text{R} \bigr\rangle = \int_X |U|^2 \, \varphi^{}_\text{L} \wedge \overline{\varphi}_\text{R}\,.
\ee
In terms of \eqref{eq:phiLR}, the closed string amplitude \eqref{eq:Mcffu} becomes
\be \label{eq:Mcre}
    \CM_\text{c} (1, \cdots, \CN) = C(1, \cdots, \CN) \, \bigl\langle \varphi^{}_\text{L}\,, \overline{\varphi^{}}_\text{R} \bigr\rangle\,,
\ee
Note that this is different from the pairing of two $k$-th twisted cohomology groups that gives rise to the intersection number of $k$-th twisted cocycles, which requires a different choice of the dual quantities, namely with $\omega^\vee = - \omega$ and $U^\vee = U^{-1}$ \cite{matsumoto1998intersection}. The duality that we are interested in here, which maps $\omega$ and $U$ to their conjugations, has been studied in \cite{mimachi2003intersection, Mizera:2017cqs}.

\subsection{Loaded Cycles and Open String Amplitudes}

Now we move on to the open string side. In terms of the Koba-Nielsen factor $U$ in \eqref{eq:KN} and the differential form $\varphi_\text{L}$ in \eqref{eq:phiL}, the open string amplitude $\CM_\text{L}$ introduced in \eqref{eq:gosa} can be cast in the form
\begin{align} \label{eq:CMLbeta}
    \Theta \bigl(\beta(1, \cdots, \CN) \, \big| \, 1, \cdots, \CN \bigr) \, \CM_\text{L} ( \beta )
    = \int_{\Delta(\beta)} U_\Delta \bigl( \beta \bigr) \, \varphi_\text{L} (z_2, \cdots, z_{\CN-2})\,,
\end{align}
where permutation $\beta$ fixes the cyclic ordering $\bigl( \beta(1), \cdots, \beta(\CN) \bigr)$\,. We introduced the $\Theta$-factor,
\be
    \Theta \bigl(i_1\,, \cdots, i_\CN \, \big| \, j_1\,, \cdots, j_{\CN} \bigr)\,,
\ee
that contributes a factor $(-1)^{n_i w_j + n_j w_i}$ if the ordering of $i$ and $j$ is the opposite in the ordered sets $(i_1\,, \cdots, i_\CN)$ and $(j_1\,, \cdots, j_{\CN})$\,.
In the case when 
\be \label{eq:betaordering}
    \bigl( \beta(1), \cdots, \beta(\CN) \bigr) = \bigl( 1, \, \sigma(2), \cdots, \sigma(\CN-2), \, \CN-1, \, \CN-2 \bigr)\,, 
\ee
the open string amplitude in \eqref{eq:CMLbeta} becomes \eqref{eq:gosa}. To be concrete, we explain different ingredients in \eqref{eq:CMLbeta} using the example \eqref{eq:betaordering}. The integration domain,
\be \label{eq:simplex}
    \Delta(1, \, \sigma(2), \, \cdots, \sigma(\CN-2), \, \CN-1, \, \CN-2)\,,
\ee
represents the $(\CN-3)$-dimensional region $0 < z_{\sigma(2)} < \cdots < z_{\sigma(\CN-2)} < 1$\,. The closure of this integration domain is an $(\CN-3)$-simplex in a smooth triangulation of the real section of the moduli space $X$\, which we refer to as $X (\mathbb{R})$\,. Since there exist more than two coalescing punctures, a minimal blowup of the moduli space is required \cite{mimachi2003intersection}. The blowup of a simplex is an associahedron $K_{\CN-1}$ \cite{tamari1954monoides, stasheff1963homotopy}. The multi-valued Koba-Nielsen factor $U$ has its branch fixed with respect to the simplex $\Delta(\beta)$ in \eqref{eq:CMLbeta}, giving rise to quantity $U_\Delta$ that is single-valued in $\Delta(\beta)$\,. In the concrete example \eqref{eq:betaordering}, we have 
\begin{align}
\begin{split}
    & \quad U_\Delta \bigl(1, \sigma(2), \cdots, \sigma(\CN-2), \, \CN-1, \, \CN \bigr) \\[2pt]
    & = \prod_{i=2}^{\CN-2} z_{\sigma(i)}^{\frac{1}{2} \alpha' K_{\text{R}1} \cdot K_{\text{R}\sigma(i)}} \bigl( 1 - z_{\sigma(i)} \bigr)^{\frac{1}{2} \alpha' K_{\text{R}(\CN-1)} \cdot K_{\text{R}\sigma(i)}} \! \prod_{\substack{k, \, \ell = 2 \\ k < \ell}}^{\CN-2} \! \bigl( z^{}_{\sigma(\ell)} - z^{}_{\sigma(k)} \bigr)^{\frac{1}{2} \alpha' K_{\text{R}\sigma(k)} \cdot K_{\text{R}\sigma(\ell)}}.
\end{split}
\end{align}
In general, for a given permutation $\beta$ in $\Delta(\beta)$\,, there corresponds a unique associahedron, for which a particular branch of the Koba-Nielsen factor is chosen \cite{Mizera:2017cqs}. 

The above pairing between the simplex $\Delta(\beta)$ and the single-valued branch choice $U (\beta)$ motivates the definition of a \textit{loaded simplex} \cite{kita1994intersection}, $\Gamma(\beta) = \Delta (\beta) \otimes U (\beta)$\,. Using the loaded simplices we define the $(\CN-3)$-th twisted homology group $H_{\CN-3} (X, \mathcal{L}_{\omega})$\,, with coefficients in the sheaf $\CL_{\omega}$ consisting of local solutions to $d\psi = \omega \, \psi$\,, where $\psi$ is a locally holomorphic function that has the formal solution $\psi \propto U$ \cite{matsumoto1998intersection, aomoto:2011theory}. Then, the loading $U$ in a loaded cycle $\Delta \otimes U$ is a section of $\CL_{\omega}$ on $\Delta$\,. The open string amplitude \eqref{eq:CMLbeta} is now defined as a pairing between the twisted homology group $H_{\CN-3} (X, \CL_{\omega})$ and the twisted cohomology group $H^{\CN-3} (X, \!\nabla_{\!\omega})$\,,
\be
    \langle \Delta(\beta) \otimes U (\beta)\,, \varphi^{}_\text{L} \rangle = \int_{\Delta (\beta)} U_\Delta (\beta) \, \varphi^{}_\text{L}\,.
\ee
Here, $\Delta(\beta) \otimes U (\beta)$ is a loaded cycle\,\footnote{We follow \cite{yoshida2013hypergeometric} and use the terminology ``loaded cycle" instead of the usual name ``twisted cycle."} that is an element of $H_{\CN-3} (X,\, \CL_{\omega})$\,, defined on the non-compact manifold $X$\,. 

Furthermore, we define the locally finite homology group $H^\text{l.f.}_{\CN-3} (X, \, \CL_{\overline{\omega}})$ \cite{kita1994intersection},
and introduce a pairing between $H^\text{l.f.}_{\CN-3} (X, \, \CL_{\overline{\omega}})$ and the cohomology group $H_\text{c}^{\CN-3} (X, \!\nabla_{\!\overline{\omega}})$ with a compact support. This pairing represents the open string amplitude $\CM_\text{R}$\,, with
\be \label{eq:CMRbeta}
     \langle \Delta(\beta) \otimes \overline{U} (\beta)\,, \overline{\varphi}^{}_\text{R} \rangle = \int_{\Delta (\beta)} \overline{U}_\Delta (\beta) \, \overline{\varphi}^{}_\text{R} = \CM_\text{R} (\beta)\,,
\ee
where $\Delta(\beta) \otimes \overline{U}^{} (\beta)$ is a loaded cycle in $H^\text{l.f.}_{\CN-3} (X, \, \CL_{\overline{\omega}})$\,.

\subsection{Intersection Number and Twisted Period Relation}

\begin{figure}
\centering
\begin{tikzpicture}
    \draw[line width = .3mm,decoration={markings, mark=at position 0.97 with {\arrow{>}}}, postaction={decorate}] (0,0) circle (.75);
    \draw (0,0) node{\scalebox{0.8}{$\bullet$}};
    \draw[white, style=thick,fill = white] (.75,-0.08) circle (.04);
    \draw[line width = .3mm,decoration={markings, mark=at position 0.475 with {\arrow{>}}}, postaction={decorate}] (4.25,0) circle (.75);
    \draw[white, style=thick,fill = white] (3.51,0.08) circle (.04);
    \draw[line width = .3mm, decoration={markings, mark=at position .5 with {\arrow{>}}}, postaction={decorate}] (.75,0)--(3.51,0);
    \draw (4.25,0) node{\scalebox{0.8}{$\bullet$}};
    \draw (4.25,-.2) node {\scalebox{0.8}{$z=1$}};
    \draw[black, scale=.01] (0,-25) node {\scalebox{0.8}{$z=0$}};
    \draw (-1,-.6) node {\scalebox{0.8}{$S_\epsilon^1 (0)$}};
    \draw (5.3,-.6) node {\scalebox{0.8}{$S_\epsilon^1 (1)$}};
    \draw (0.9,0.2) node {\scalebox{0.8}{$\epsilon$}};
    \draw (3.02,0.2) node {\scalebox{0.8}{$1-\epsilon$}};
\end{tikzpicture}
\caption{\small Regularization of the path $[0,1]$\,. $S_\epsilon^1 (z)$ is a circle centered at $z$ with radius $\epsilon$\,.} 
\label{fig:pochhammer}
\end{figure}
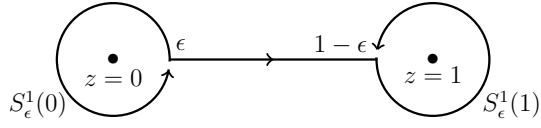

Finally, we consider a pairing between $H_{\CN-3} (X, \, \CL_{{\omega}})$ and the locally finite homology group $H^{\text{l.f.}}_{\CN-3} (X, \, \CL_{\overline{\omega}})$\,. This pairing gives rise to a geometric interpretation for the inverse KLT kernel as an intersection number between two loaded cycles \cite{Mizera:2017cqs}. There exists an isomorphism map \cite{aomoto:2011theory},
\be
    \text{reg\,: } H_{\CN-3}^\text{l.f.} (X, \CL_{\overline{\omega}}) \rightarrow H_{\CN-3} (X, \CL^{}_{\overline{\omega}^{}})\,.
\ee
The map ``reg" can be realized by a regularization of the loaded cycles. For example, when $\CN=4$\,, and when the tachyonic vertex operators are concerned, we encounter the integral 
\be \label{eq:integralex}
    \int_{0 < z_2 < 1} U(1, 2, 3, 4) \, \varphi_\text{L} (z_2)\,,
\ee
with
\be \label{eq:Uphi}
    U(1, 2, 3, 4) = z^{\tfrac{1}{2} \alpha' K_{\text{R}1} \cdot K_{\text{R}2}} \, (1-z)^{\tfrac{1}{2} \alpha' K_{\text{L}2} \cdot K_{\text{R}3}},
        \qquad%
    \varphi_\text{L} (z_2) = f_\text{L} (z) \, dz\,.
\ee
This integral is equal to the open string amplitude \eqref{eq:ILR1} up to the $\Theta$-factor as specified in \eqref{eq:CMLbeta}.
Moreover, \eqref{eq:integralex} only converges when both the exponents (that we have assumed to be non-integers) in \eqref{eq:Uphi} are larger than $-1$. The regularization of the integration domain $(0,1)$ is the Pochhammer contour, which gives rise to the regularization $\Delta (1, 2, 3, 4)$ of the path $[0,1]$ as in Figure \ref{fig:pochhammer}. We have 
\begin{subequations}
\begin{align}
    \Gamma (1, 2, 3, 4) & = [0,1] \otimes U(1, 2, 3, 4) \in H_1^\text{l.f.} (X, \CL_{\omega})\,, \\[2pt]
    \text{reg } \Gamma (1, 2, 3, 4) & = \Delta (1, 2, 3, 4) \otimes U (1, 2, 3, 4) \in H_1 (X, \, \CL_{{\omega}})\,,
\end{align}
\end{subequations}
which can be generalized to arbitrary $\CN$\,. See \cite{mimachi2003intersection, aomoto:2011theory, Mizera:2017cqs} for further details.

\begin{figure}
\centering
    \begin{subfigure}[b]{1\textwidth}
    \centering
    \begin{tikzpicture}
        \draw (6.5,0) node {$\phantom{\ldots}$};
        \draw (6.5,-.3) node {\scalebox{.8}{$\phantom{z= \infty}$}};
        \draw[white,line width = .3mm, decoration={markings, mark=at position 0.6 with {\arrow{>}}}, postaction={decorate}] (4.25,0)--(6,0);
        \draw[darkred, line width = .3mm, decoration={markings, mark=at position 1 with {\arrow{>}}}, postaction={decorate}] (-2,0)--(-0.12,0);
        \draw (0,0) node{\scalebox{.8}{$\bullet$ }};
        \draw[line width = .3mm, decoration={markings, mark=at position 0.97 with {\arrow{>}}}, postaction={decorate}, darkblue] (0,0) circle (.75);
        \draw[white, style=thick,fill = white] (.75,-0.08) circle (.04);
        \draw[line width = .3mm, decoration={markings, mark=at position .5 with {\arrow{>}}}, postaction={decorate}, darkblue] (.75,0)--(3.51,0);
        \draw[scale=.01] (0,-25) node {\scalebox{0.8}{$z=0$}};
        \draw (-2.5,0) node {$\ldots$};
        \draw (0.9,0.2) node {\scalebox{0.8}{$\epsilon$}};
        \draw[scale=.01] (-250,-25) node {\scalebox{0.8}{$z=-\infty$}};
        \draw[line width = .3mm,decoration={markings, mark=at position 0.475 with {\arrow{>}}}, postaction={decorate}, darkblue] (4.25,0) circle (.75);
        \draw[white, style=thick,fill = white] (3.51,0.08) circle (.04);
        \draw (3.02,0.2) node {\scalebox{0.8}{$1-\epsilon$}};
        \draw (4.25,0) node{\scalebox{0.8}{$\bullet$}};
        \draw (4.25,-.2) node {\scalebox{0.8}{$z=1$}};
        \draw (-1.02,0.2) node {\scalebox{0.8}{$-\epsilon$}};
    \end{tikzpicture}
    \caption{\small Intersection between $\text{reg}\,\Gamma (1, 2, 3, 4)$ (in blue) and $\Gamma(2, 1, 3, 4)$ (in red) at $z = - \epsilon$\,. Using \eqref{eq:in}, the intersection number is computed to be $\bigl[ 2 \, i \sin \bigl( \frac{1}{2} \, \pi \alpha' K_{\text{R}1} \! \cdot \! K_{\text{R}2} \bigr) \bigr]^{-1}$, whose inverse matches the KLT kernel in \eqref{eq:KLT1}.\\} 
    \label{IntersectionLeft}
\end{subfigure}
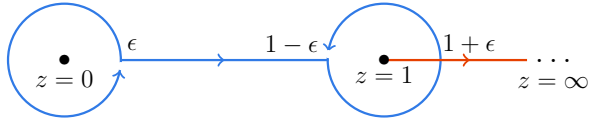
\begin{subfigure}[b]{1\textwidth}
    \centering
    \begin{tikzpicture}
        \draw (-2.5,0) node {$\phantom{\ldots}$};
        \draw[scale=.01] (-250,-25) node {\scalebox{0.8}{$\phantom{z=-\infty}$}};
        \draw[white, line width = .3mm, decoration={markings, mark=at position 1 with {\arrow{>}}}, postaction={decorate}] (-2,0)--(-0.12,0);
        \draw[line width = .3mm, decoration={markings, mark=at position 0.97 with {\arrow{>}}}, postaction={decorate}, darkblue] (0,0) circle (.75);
        \draw[white, style=thick,fill = white] (.75,-0.08) circle (.04);
        \draw[line width = .3mm, decoration={markings, mark=at position .5 with {\arrow{>}}}, postaction={decorate}, darkblue] (.75,0)--(3.51,0);
        \draw[black, scale=.01] (0,-25) node {\scalebox{0.8}{$z=0$}};
        \draw (0.9,0.2) node {\scalebox{0.8}{$\epsilon$}};
        \draw[line width = .3mm,decoration={markings, mark=at position 0.475 with {\arrow{>}}}, postaction={decorate}, darkblue] (4.25,0) circle (.75);
        \draw[white, style=thick,fill = white] (3.51,0.08) circle (.04);
        \draw (3.02,0.2) node {\scalebox{0.8}{$1-\epsilon$}};
        \draw (4.25,-.2) node {\scalebox{0.8}{$z=1$}};
        \draw[darkred,line width = .3mm, decoration={markings, mark=at position 0.6 with {\arrow{>}}}, postaction={decorate}] (4.25,0)--(6.15,0);
        \draw (6.5,0) node {$\ldots$};
        \draw (0,0) node {\scalebox{0.8}{$\bullet$}};
        \draw (6.5,-.3) node {\scalebox{.8}{$z= \infty $}};
        \draw (4.25,0) node{\scalebox{0.8}{$\bullet$}};
        \draw (5.38,0.22) node {\scalebox{0.8}{$1+\epsilon$}};
    \end{tikzpicture}
    \caption{\small Intersection between $\text{reg}\,\Gamma (1, 2, 3, 4)$ (in blue) and $\Gamma(1, 3, 2, 4)$ (in red) at $z = 1 + \epsilon$\,. Using \eqref{eq:in}, the intersection number is computed to be $\bigl[ 2 \, i \sin \bigl( \frac{1}{2} \, \pi \alpha' K_{\text{R}2} \! \cdot \! K_{\text{R}3} \bigr) \bigr]^{-1}$, whose inverse matches the KLT kernel in \eqref{eq:KLT2}.} 
    \label{IntersectionRight}
    \end{subfigure}
\caption{\small Intersection numbers between different loaded cycles.}
\label{fig:intersection}
\end{figure}

The intersection number between the two loaded cycles is defined with respect to the intersection points between the paths $\Gamma$ and reg$\,\Gamma$\,. In general, we define the intersection number as the following pairing between $\text{reg}\,\Gamma(\beta) \in H^\text{l.f.}_{\CN-3} (X, \CL_{\omega})$ and $\Gamma(\gamma) \in H_{\CN-3}^\text{l.f.} (X, \CL_{\overline{\omega}})$ \cite{mimachi2003intersection}:
\be \label{eq:in}
    \bigl\langle \text{reg} \, \Gamma (\beta)\,, \Gamma (\gamma) \bigr\rangle = \!\!\!\!\! \sum_{p \, \in \, \Delta \, \cap \, \Delta'} \!\!\!\! \text{I}_p (\Delta\,, \Delta') \, \frac{U_\Delta \, \overline{U}_{\!\Delta'}}{|U|^2} \bigg|_{p}\,,
\ee
where we defined 
\be
    \text{reg} \, \Gamma (\beta) = \Delta \otimes U_{\Delta}\,,
        \qquad%
    \Gamma (\gamma) = \Delta' \otimes \overline{U}_{\Delta'}\,.
\ee
Here, $\Delta$ and $\Delta'$ are $\CN-3$ simplices and I${}_p$ denotes the topological intersection number at the intersecting point $p$\,, which only depends on the relative orientations of $\Delta$ and $\Delta'$ at $p$\,. It is shown in \cite{Mizera:2017cqs} that \eqref{eq:in} precisely gives the inverse of the KLT kernel that is a function of $K_{\text{R}i}$\,. For example, see Figure \ref{fig:intersection} for intersection numbers that give rise to the inverse kernels in the KLT relations from \eqref{eq:Mc1234}. See \cite{Mizera:2017cqs} for detailed calculation. This intersection number between two loaded cycles also receives the physical interpretation as the $\alpha'$-corrected bi-adjoint scalar amplitudes \cite{Mizera:2016jhj}. 

Assembling the ingredients introduced above, we now rewrite the KLT relation for winding strings as a twisted Riemann's period relation,
\begin{align} \label{eq:Rpr}
    \bigl\langle \varphi^{}_\text{L}, \, \overline{\varphi}^{}_\text{R} \bigr\rangle = \sum_{\Gamma' (\beta), \, \Gamma (\gamma)} \!\! \bigl\langle \varphi^{}_\text{L}\,, \Gamma' (\beta) \bigr\rangle \, m^{-1}(\beta|\gamma) \, \bigl\langle \Gamma (\gamma)\,, \overline{\varphi}^{}_\text{R} \bigr\rangle\,,
\end{align}
where $m^{-1}(\beta|\gamma)$ is the inverse of the intersection matrix $m(\beta|\gamma) \equiv \bigl\langle \Gamma' (\beta)\,, \Gamma (\gamma) \bigr\rangle$\,.
Also note that 
\begin{subequations}
\begin{align}
    \varphi^{}_\text{L} & \in H^{\CN-3} (X, \nabla_{\!\omega})\,, 
        &%
    \Gamma' & \in H_{\CN-3} (X, \CL_\omega)\,, \\[2pt]
    \varphi^{}_\text{R} & \in H^{\CN-3}_\text{c} (X, \nabla_{\!\overline{\omega}})\,,
        &%
    \Gamma & \in H^\text{l.f.}_{\CN-3} (X, \CL_{\overline{\omega}})\,.
\end{align}
\end{subequations}
Here, the sum over $\Gamma'$ and $\Gamma$ are taken over generic bases of the loaded cycles in $H_{\CN-3} (X, \CL_{\omega})$ and $H^\text{l.f.}_{\CN-3} (X, \CL_{\overline{\omega}})$\,, respectively. In the case of zero winding, \eqref{eq:Rpr} reduces to the one discussed in \cite{mimachi2003intersection, Mizera:2017cqs}.
Taking the relations to closed and open string amplitudes \eqref{eq:Mcre}, \eqref{eq:CMLbeta}, and \eqref{eq:CMRbeta} into account, we find that \eqref{eq:Rpr} becomes
\be \label{eq:itKLT}
    \CM_\text{c} = \sum_{\beta, \, \gamma} C(\beta) \, \CM_\text{L} (\beta) \, m^{-1}(\beta|\gamma) \, \CM_\text{R} (\gamma)\,,
\ee
which succinctly summarizes the KLT relations given in \eqref{eq:Mcresult}, \eqref{eq:Mcresult1N-2}, and \eqref{eq:Mcresult1N-22}.\,\footnote{The KLT relation \eqref{eq:itKLT} also include the type with $\text{reg} \, \Gamma (\beta) = \Gamma' (\beta)$\,. In the case of $\CN=4$\,, this involves the intersection between $\text{reg} \, \Gamma(1, 2, 3, 4)$ and $\Gamma (1, 2, 3, 4)$\,, instead of $\Gamma (2, 1, 3, 4)$ or $\Gamma (1, 3, 2, 4)$ in Figure \ref{fig:intersection}. In general, such KLT relations can be obtained from the KLT relations discussed in the bulk of the paper by using the stringy Kleiss-Kuijf and BCJ relations in \cite{BjerrumBohr:2009rd, Stieberger:2009hq}.} If we start with the multi-valued function, 
\be
    U (z_2, \cdots, z_{\CN-2}) = \prod_{\substack{i, \, j = 1 \\[2pt] i < j}}^{\CN-1} z^{\frac{1}{2} \alpha' K_{\text{L}i} \cdot K_{\text{L}j}}_{ji},
\ee
instead of the expression for $U$ in \eqref{eq:KN} that has been considered through this appendix, we will land on the twisted Riemann's period relation that represents the same KLT relations but with ``L" and ``R" swapped (which of course swaps $K_{\text{L}i}$ and $K_{\text{R}i}$). 

\newpage

\bibliographystyle{JHEP}
\bibliography{KLT}

\end{document}